\documentclass[12pt,a4paper]{article}
\usepackage[latin1]{inputenc}
\usepackage{amsmath}
\usepackage{amsfonts}
\usepackage{amssymb}
\usepackage[titletoc,title]{appendix}
\usepackage{geometry}
\usepackage[hidelinks]{hyperref}
\begin{document}

\title{Boundary Effects on the Thermodynamics of Quantum Fields Near a Static Black Hole}

\author{Emine Ertu\u{g}rul\thanks{Email: emine.ertugrul@bogazici.edu.tr}, Birses Debir\thanks{Email: birses.debir@gmail.com}, Levent Akant\thanks{Email: levent.akant@bogazici.edu.tr},}
\date{\small Department of Physics, Bo\u{g}azi\c{c}i University, 34342 Bebek, Istanbul, Turkey}

\maketitle
\begin{abstract}
We investigate thermodynamics of a non-interacting quantum field in a static black hole background. The horizon divergences are regulated by the brick wall method, which consists of subjecting the quantum field to Dirichlet boundary conditions on a surface (the brick wall) just outside the horizon. Using heat kernel and Mellin transform methods, we derive high-temperature expansions for the free energy and entropy and study the boundary and higher-order geometric effects on the horizon divergences induced by the brick wall. We consider real scalar, complex scalar, and Dirac fields in Schwarzschild, Reissner-Nordstr\"{o}m and dilatonic black hole backgrounds, as well as in their near-horizon geometries. By evaluating the high-temperature expansion of the entropy (up to a certain order) at the Hawking temperature, we show that, for a given field type, the leading horizon divergence is the same for all the metrics considered. Moreover, we show that the different orders in the high-temperature expansion become comparable at the Hawking temperature and compare our findings with existing results in the literature. We derive an explicit formula for the sub-leading horizon divergence expressed in terms of the field mass, surface gravity, horizon area, and dilaton parameter, which is applicable to all the exact metrics considered. We also consider the possibility of using Neumann boundary conditions to regularize the horizon divergences.

\end{abstract}

\maketitle

\section{Introduction}

The analysis of thermodynamic properties of a quantum field around a black hole is an essential step towards a deeper understanding of the quantum mechanical properties of black holes \cite{Hawking1, Hawking2, Bekenstein}.  It is well known \cite{tHooft, Davies, Vega, Susskind, Kabat, Solodukhin, Barbon, Emperan, Alwis0, Alwis, Solodukhin1, Fursaev, Belgiorno} that the free energy and the entropy of a quantum field around a black hole are plagued by divergences caused by the presence of the horizon. Therefore, one needs to devise regularization methods to tackle these so-called horizon divergences. One such regularization procedure is 't Hooft's brick wall method \cite{tHooft} which consists of confining the quantum field inside a spherical shell $\mathcal{S}$ around a static black hole bounded by two Dirichlet walls, \textit{i.e.} two hypersurfaces where the quantum fields are subject to Dirichlet boundary conditions. The inner Dirichlet wall (the brick wall), situated just outside the black hole horizon, regulates the horizon divergence; and the outer wall provides an infrared cutoff. Our aim in this paper is to investigate the effects of the brick wall on the thermodynamics of real scalar, complex scalar and Dirac fields in Schwarzschild, Reissner-Nordstr\"{o}m (RN), dilatonic, and approximate near-horizon metric backgrounds.
We will derive a high-temperature expansion for the free energy and the entropy in aforementioned static black hole backgrounds in terms of appropriate heat kernel coefficients using the methods developed in \cite{Akant}. In our explicit calculations we will consider the first four orders of the expansion. 

As a result, we will show that the entropy is given as
 \begin{equation}
    S=B\, \frac{A_{H}}{\delta^{2}}+C \log\frac{\kappa\delta^{2}}{2(r_2-r_H)},
  \end{equation}
where $A_{H}$ is the area of the event horizon and $\delta$ is the proper radial distance between the horizon and the brick wall. The coefficients $B$ and $C$ will be derived from the high-temperature expansion and will involve boundary and geometric contributions due to the brick wall. This form of entropy is in agreement with the results of \cite{Solodukhin, Alwis0, Alwis} where the coefficients $B$ and $C$ were calculated in the high-temperature limit. Here, we will go beyond the limiting behavior, derive the high-temperature expansion and use it to study the effects of the brick wall on the entropy. It will be shown that when the entropy is evaluated at the Hawking temperature $T_{H}$, with explicit expressions for the heat kernel coefficients, the different terms of the expansion become comparable, at least up to the order considered. Consequently, the coefficient $B$ will be seen to consist, apart from a logarithmic term in surface gravity $\kappa$ encountered in the case of scalar field, of a numerical series that depends only on the type of the quantum field. In Sec.3 we will make further remarks on this logarithmic term after its origin is clarified in Sec.2 as a general feature of the high-temperature expansion for Bose-Einstein statistics  \cite{Akant,Dowker1,Actor,Dowker2,Dowker3,Kirsten1,Kirsten2,T1,T2}. In particular, we will comment on the possibility of it becoming a purely numerical factor by the inclusion of higher-order terms of the high-temperature expansion. It will be shown that such a term is absent in the fermionic case.

In \cite{Ordonez1} the enhancement of the leading horizon divergence $B$ was shown for a scalar field in the near-horizon geometry by a scaling analysis of the high-temperature expansion, which, however, does not yield the explicit values of the terms appearing in $B$. In addition, it was argued that this is a holographic effect arising from the conformal invariance of the Klein-Gordon equation in the near-horizon geometry (conformal quantum mechanics) \cite{Ordonez2, Ordonez2.5, Ordonez3}. In the near-horizon geometry we will derive a closed-form expression for the entropy which will allow us to prove the enhancement property at $T_{H}$ without regard to the high-temperature expansion. We will also show that the coefficient $B$ is the same in both the exact and near-horizon geometries, apart from the aforementioned logarithmic term in $\kappa$ which, as we will show, becomes a numerical factor in the near-horizon geometry. In \cite{Ordonez1} it was conjectured that the high-temperature expansion of the leading horizon divergence for the exact and near-horizon metrics must be identical. Our explicit calculations of the coefficient $B$ in both cases will verify this conjecture, at least to the order considered (and up to the complication due to the logarithmic term).

As for the sub-leading horizon divergence, we will derive a closed form expression for the coefficient $C$, that applies to all the exact metrics considered, and gives $C$ as a function of mass $m$ of the field, $\kappa$, $A_{H}$ and the dilaton parameter $a$. In \cite{Solodukhin, Alwis} where only the leading term of the high-temperature expansion was considered, it was shown that $C$ vanishes for the near-horizon metric. Here we will find that this result only holds at the leading order, and by examining the higher-order terms of the high-temperature expansion we will show that $C$ is, in fact, non-vanishing in the near-horizon geometry.

The total entropy is given by adding the entropy $S$ of the quantum field to the purely gravitational entropy of the black hole, which at tree level is given by the Bekenstein-Hawking entropy
\begin{equation}
S_{total}=\frac{A_{H}}{4G}+S
\end{equation}
The leading horizon divergence is then canceled by renormalizing the gravitational constant as
\begin{equation}
\frac{1}{G_{ren}}=\frac{1}{G}+\frac{4B}{\delta^2},
\end{equation}
whereas the sub-leading divergence can be canceled by curvature squared counterterms in the gravitational action \cite{Solodukhin, Solodukhin1}. Thus we see that the renormalization of $G$ involves the background independent numerical factor $B$ (apart from the aforementioned logarithmic term in surface gravity in the case of scalar field, which however becomes a numerical factor in the near-horizon geometry). On the other hand, the background dependence of $C$ beyond the leading order will lead to the background dependent renormalization of the coefficients of curvature squared counterterms in the gravitational action. We will not discuss these counterterms in this paper. 

Thermodynamics of a quantum field in the extreme RN background is ill-defined, since at the leading order of the high-temperature expansion one encounters higher-order horizon divergences \cite{Alwis0, Alwis} which cannot be renormalized by the purely gravitational entropy of the black hole. According to the proposal of references \cite{Alwis0, Alwis}, a sensible thermodynamics in the extreme background is possible if the extreme limit is taken after the thermodynamic quantities and the horizon divergences are derived in the non-extreme case. Here we will follow this strategy and show that the leading horizon divergence of the non-extremal case, which now includes higher-order boundary and geometric contributions, vanishes in this limit. In particular, this means that Newton's constant is not renormalized in the extreme background.   

Going beyond the leading order of the high-temperature expansion we encounter the effects of the Dirichlet boundary conditions via the surface terms present in the heat kernel coefficients of higher index. This prompts us to study the effects of alternative boundary conditions on the thermodynamics of the quantum field. Here, we will examine the possibility of employing Neumann boundary conditions as an alternative to the usual Dirichlet boundary conditions, though without delving into a deeper physical motivation. We will find that the leading horizon divergence, which in general gets contributions from both bulk and boundary terms of the heat kernel coefficients, is different from the result found for the Dirichlet boundary conditions. On the other hand, the sub-leading horizon divergence will be seen to remain unaltered, as expected from the fact that only the bulk terms in the heat kernel coefficients contribute to it.

For a system of non-interacting spinless bosons (\textit{e.g.} free real scalar field) at the inverse temperature $\beta$ the free energy is given by
\begin{equation}\label{freeen}
  \mathcal{F}=\frac{1}{\beta} \textrm{Tr} \log(1-e^{-\beta H}),
\end{equation}
where $H$ is the single particle Hamiltonian of the boson. Expanding the logarithm, we get
\begin{equation}
  \mathcal{F}=-\frac{1}{\beta} \sum_{k=1}^{\infty} \frac{1}{k}\textrm{Tr} e^{-k\beta H},
\end{equation}
At this point, one may try to plug the heat kernel expansion of $\textrm{Tr} e^{-k\beta H}$ into the above sum to derive the high-temperature expansion of the free energy. However, such a procedure is not a reliable expansion method due to the competition between large values of the summation index $k$ and the small value of $\beta$.
 Indeed, a rigorous analysis of the thermodynamics of the free scalar field on a curved manifold \cite{Akant, Dowker1,Actor,Dowker2,Dowker3,Kirsten1,Kirsten2,T1,T2} reveals the existence of a logarithmic term in the high-temperature expansion of the free energy that cannot be detected in the naive expansion scheme described above. The use of the naive method is the reason why such a logarithmic term in the coefficient $B$ is missing in \cite{Ordonez1}. In the present paper, the high-temperature expansion of the free energy will be derived using the method of \cite{Akant}, which exploits the role of the Mellin transform in deriving asymptotic expansions \cite{Flajolet}. In the case of scalar field this will lead to the aforementioned logarithmic term in the high-temperature expansion. However, we will see that, in contrast to the scalar case, the logarithmic term is absent for the Dirac field.

Instead of the original metric, we will work with the optical metric, which is related to the former by a conformal transformation. The importance of the optical metric in the path integral analysis of black hole thermodynamics is well known \cite{gibbons, Barbon, Alwis}. Here we will prefer to work directly with the field equations. The main technical advantage of the optical metric is that it allows us to express the single particle energies as eigenvalues of a Hamiltonian constructed therefrom. This, in turn, allows us to study thermodynamic quantities using the heat kernel and Mellin transform techniques \cite{Akant}. Here we also note the importance of the brick wall; it constrains the system in a region where the conformal transformation is free of singularities.

The brick wall method should not be confused with the often used volume cut-off method in which one works within the same spherical shell used in the brick wall method, but without imposing any boundary conditions (see \cite {FrolovFursaev} for a comparison of various techniques employed in the calculation of the thermodynamic entropy of quantum fields near a black hole). In particular, the heat kernel expansion used in the volume cut-off method does not involve the boundary terms. Thus, in light of our results, we conclude that the volume cut-off method cannot be equivalent to the brick wall method which contains boundary contributions to the heat kernel.

The leading order contribution to the entropy in the high-temperature expansion was first calculated in \cite{Solodukhin}  by a rather involved analysis based on a Liouville type quantum field theory. Later, this result was reproduced in \cite{Barbon, Alwis} by a simpler analysis based on the use of the optical metric. Our leading order result will be in complete agreement with the findings of the above papers. Higher-order corrections in the expansion were obtained in the volume cut-off method by \cite{ Vanzo1, Vanzo2}. The analysis of the problem in near-horizon geometry was given in \cite{Ordonez1}. For a WKB analysis of the problem see \cite{Ghosh1, Ghosh2, Ghosh3, Ghosh4, Ordonez2, Ordonez2.5, Ordonez3, Sarkar, Kim}. The effect of boundaries in particle detection in AdS space was discussed recently in \cite{Emelyanov}. For a different approach to the thermodynamics of a quantum field in a black hole background see \cite{FrolovNovikov, Barvinsky}. For an extensive review of various regularization methods employed in entropy calculations around a black hole, see \cite{Solodukhin2}.

Here is a brief outline of this paper. In Sec.2 we will briefly review the optical metric and derive the high-temperature expansions of the free energy and the entropy, following the methods developed in \cite{Akant}. We will first express the free energy as a harmonic sum and then apply the Mellin transform method together with the heat kernel expansion to derive the expansions for real scalar, complex scalar and Dirac fields. In Sec.3, we will apply the results of Sec.2 to Schwarzschild, RN, and the dilatonic black hole backgrounds and derive the boundary and geometric effects on the horizon divergences. We will also discuss the aforementioned logarithmic term and the case of extremal RN background. In Sec.4 we will consider thermodynamics in the near-horizon geometry, derive a closed formula for the entropy, give a concise proof of the enhancement of entropy at the Hawking temperature, and compare the near-horizon result to those obtained for exact metrics. In Sec.5, we will examine the possibility of using Neumann boundary conditions as regulators of horizon divergences. In the Appendix, we will present the explicit calculations leading to the results of Sec.3.

\section{Thermodynamics Around the Black Hole}

In this section, we will derive the high-temperature expansion for scalar and Dirac fields in the static background
\begin{equation}
  ds^{2}=-F dt^{2}+g_{ij}dx^{i}dx^{j}.
\end{equation}
We will start our derivation in a $d+1$ dimensional space-time, but at an appropriate stage we will set $d=3$. The quantum field will be confined within a region $\mathcal{S}$ and will be assumed to vanish at its boundary $\partial \mathcal{S}$.

\subsection{Free Scalar Field}\label{bosegas}
Ignoring the zero temperature contribution, which has no effect on the entropy, the free energy of a complex scalar can be written as $\mathcal{F}_{cs}(\beta,\mu)=\mathcal{F}(\beta,\mu)+\mathcal{F}(\beta,-\mu)$. Here $\beta$ is the inverse temperature, $\mu$ is the chemical potential, and
\begin{eqnarray}
\mathcal{F}(\beta,\mu) = \frac{1}{\beta}\sum_{\sigma}\log\left[1-e^{-\beta(\epsilon_{\sigma}-\mu)}\right],
\end{eqnarray}
where $\epsilon_{\sigma}$'s are the single-particle energies determined by solving the Dirichlet problem for the Klein-Gordon equation coupled to the space-time scalar curvature $\mathcal{R}$
\begin{equation}
  \left(-\Box+\xi \mathcal{R}+m^{2}\right)f_{\sigma}=0,
\end{equation}
by the separation of variable $f_{\sigma}(t,x)=e^{-i\epsilon_{\sigma}t}\phi_{\sigma}(x)$ \cite{Thorne}.

Note that for $\mu=0$, $\mathcal{F}(\beta,0)$ is the free energy $\mathcal{F}_{rs}(\beta)$ of the real scalar field and
\begin{equation}\label{fss}
  \mathcal{F}_{cs}(\beta,0)=2\mathcal{F}_{rs}(\beta).
\end{equation}
In what follows, we will not indicate the $\beta$ and $\mu$ dependencies of $\mathcal{F}$ explicitly in our notation unless they are needed to be emphasized.

Under the conformal scaling transformation
\begin{equation}
  \overline{g}_{\mu\nu}=\Omega^{2}g_{\mu\nu},\;\;\;\overline{f}_{\sigma}=\Omega^{\frac{1-d}{2}}f_{\sigma},
\end{equation}
 with $\Omega > 0$, the KG equation becomes
\begin{equation}
   \left\{-\Box_{c}+\frac{d-1}{4d}R+\Omega^{-2}\left[m^{2}+\left(\xi-\frac{d-1}{4d}\right)\mathcal{R}\right]\right\}\overline{f}_{\sigma}=0,
\end{equation}
where $\Box_{c}$ and $R$ are, respectively, the D'Alembertian and the scalar curvature of the transformed metric $\overline{g}_{\mu\nu}$. In particular if we choose $\Omega=F^{-1/2}$ we get the so-called optical metric
\begin{equation}
   d\overline{s}^{2}=- dt^{2}+\overline{g}_{ij}dx^{i}dx^{j},
\end{equation}
where 
\begin{equation}
  \overline{g}_{ij}=F^{-1}g_{ij}.
\end{equation}
Thus, KG equation can be written as,
\begin{equation}
   \left\{\partial_{0}^{2}-\Delta+\frac{d-1}{4d} R+F\left[m^{2}+\left(\xi-\frac{d-1}{4d}\right)\mathcal{R}\right]\right\}\overline{f}_{\sigma}=0.
\end{equation}
Here $\Delta$ is the Laplacian of the metric $\overline{g}_{ij}$. Since $F$ does not depend on $t$ we can still write the solution in the separated form $\overline{f}_{\sigma}(t,x)=e^{-i\epsilon_{\sigma}t}\overline{\phi}_{\sigma}(x)$. Thus, we arrive at the eigenvalue problem 
\begin{equation}
   \left\{-\Delta+\frac{d-1}{4d}R+F\left[m^{2}+\left(\xi-\frac{d-1}{4d}\right)\mathcal{R}\right]\right\}\overline{\phi}_{\sigma}=
   \epsilon_{\sigma}^{2}\overline{\phi}_{\sigma}(x).
\end{equation}
Defining 
\begin{equation}
  U_{1}=\frac{d-1}{4d}R+F\left[m^{2}+\left(\xi-\frac{d-1}{4d}\right)\mathcal{R}\right],
\end{equation}
we get 
\begin{equation}
  H_{1}\overline{\phi}_{\sigma}=\epsilon_{\sigma}^{2}\overline{\phi}_{\sigma}(x),
\end{equation}
with
\begin{equation}
H_{1}=-\Delta+U_{1}.
\end{equation}

Now, coming back to the free energy and proceeding as in \cite{Akant} we write it as
\begin{eqnarray}
  \mathcal{F} &=& -\frac{1}{\beta}\sum_{\sigma}\sum_{k=1}^{\infty}\frac{1}{k}e^{k\beta\mu}e^{-k\beta\epsilon_{\sigma}} =  -\sum_{k=1}^{\infty}\frac{1}{k\beta}e^{k\beta\mu}\sum_{\sigma}e^{-k\beta\epsilon_{\sigma}}.
\end{eqnarray}
Using the subordination identity
\begin{eqnarray}\label{subor}
  e^{-b\sqrt{x}} =
  \frac{b}{2\sqrt{\pi}}\int_{0}^{\infty}\frac{du}{u^{3/2}}\,e^{-\frac{b^{2}}{4u}}\,e^{-ux},
\end{eqnarray}
we get
\begin{equation}\label{mp}
\sum_{\sigma}e^{-k\beta\epsilon_{\sigma}}=\sum_{\sigma}e^{-k\beta m'(m'^{-1}\epsilon_{\sigma})}=\frac{k\beta m'}{2\sqrt{\pi}}\int_{0}^{\infty}\frac{du}{u^{3/2}}e^{-\frac{(k\beta m')^{2}}{4u}-u}\,\text{Tr}\,e^{-um'^{-2}H},
\end{equation}
and
\begin{eqnarray}\label{F}
   \mathcal{F}=  -\sum_{k=1}^{\infty}e^{k\beta\mu}\frac{m'}{2\sqrt{\pi}}\int_{0}^{\infty}\frac{du}{u^{3/2}}e^{-\frac{(k\beta m')^{2}}{4u}-u}\, \text{Tr}\,e^{-um'^{-2}H}.
\end{eqnarray}
Here $m'$ is a mass parameter that will be used to form the dimensionless expansion parameter $\beta m'$. Here we also defined
\begin{equation}
  H \equiv H_{1}-m'^{2}=-\Delta+U,
\end{equation}
with a shifted potential energy
\begin{equation}\label{ushift}
  U\equiv U_{1}-m^{'2}=\frac{d-1}{4d}R+F\left[m^{2}+\left(\xi-\frac{d-1}{4d}\right)\mathcal{R}\right]-m^{'2}.
\end{equation}
Defining $y=\beta m'$, let us rewrite the free energy as
\begin{equation}\label{harm}
  \mathcal{F}(y)=-m'\sum_{k=1}^{\infty}e^{ky\overline{\mu}_{r}}G(ky),
\end{equation}
with
\begin{equation}\label{Gdef}
  G(y)=\frac{e^{y\overline{\epsilon}_{0}}}{2\sqrt{\pi}}\int_{0}^{\infty}\frac{du}{u^{3/2}}e^{-\frac{y^{2}}{4u}-u} \,\text{Tr}\,e^{-u(m')^{-2}H}.
\end{equation}
and
\begin{equation}\label{mur}
  \mu_{r}=\mu-\epsilon_{0},\;\;\;\;\;\;\overline{\mu}=\frac{\mu}{m'},\;\;\;\;\;\;\overline{\mu}_{r}=\frac{\mu_{r}}{m'},\;\;\;\;\;\;\overline{\epsilon}_{0}=\frac{\epsilon_{0}}{m'}.
\end{equation}
The sum in (\ref{harm}) is a harmonic sum which, by definition, is a series of the form
\begin{equation}\label{harmsum}
  \sum_{k=1}^{\infty}d_{k}F(ky),
\end{equation}
where $d_k$'s are numerical coefficients and $F(y)$ is a function of $y$. The small $y$ asymptotic expansion of this harmonic sum is  determined by the poles and residues of the meromorphic extension of the Mellin transform of the harmonic sum \cite{Flajolet}.

The Mellin transform of a function $f(y)$ is given by
\begin{equation}
 \widetilde{f}(s)=(\mathcal{M}f(y))(s)= \int_{0}^{\infty}dy\, y^{s-1}f(y).
\end{equation}
Let us also recall the definition of the Laplace-Mellin transform \cite{JorLang} of $f$
\begin{equation}
  \widetilde{f}(s,z)=(\mathcal{LM}f(y))(s,z)=\int_{0}^{\infty}dy\, y^{s-1}e^{-zy}f(y).
  \end{equation}
The asymptotic expansion of $f(y)$ is then given according to the following rule \cite{Flajolet}. If the Mellin transform has the singular expansion ($\asymp$ means the singular part of the expansion),
\begin{equation}
\widetilde{f}(s)\asymp\sum_{w,k}\frac{A(w,k)}{(s-w)^{k+1}},
\end{equation}
then as $y\rightarrow 0$
\begin{equation}\label{mellinasy}
 f(y)\sim \sum_{w,k}A(w,k)\frac{(-1)^{k}}{k!}y^{-w}(\log y)^{k}.
\end{equation}
If $f(y)$ is given by a harmonic sum of the form (\ref{harmsum}) then it is easy to see that
\begin{equation}
  \widetilde{f}(s)=\left(\sum_{k=1}^{\infty}\frac{d_{k}}{s^{k}}\right)\widetilde{F}(s),
\end{equation}
Now we can apply this method to $ \mathcal{F}(y)$ whose Mellin transform is 
\begin{equation}
  \mathcal{M}(\mathcal{F}(y))(s)=-m'\zeta(s)\mathcal{M}(e^{y\overline{\mu}_{r}}G(y))(s),
\end{equation}
where $\zeta(s)$ is the Riemann $\zeta$ function
\begin{equation}
  \zeta(s)=\sum_{k=1}^{\infty}\frac{1}{s^{k}}.
\end{equation}
In terms of the Laplace-Mellin transform of $G(y)$ we can write this as
\begin{equation}
  \mathcal{M}(\mathcal{F}(y))(s)=-m'\zeta(s)\widetilde{G}(s,-\overline{\mu}_{r}),
\end{equation}
where
\begin{equation}
  \widetilde{G}(s,-\overline{\mu}_{r})=\int_{0}^{\infty}dy\, y^{s-1}e^{\overline{\mu}_{r} y}G(y).
\end{equation}
The singularities of this integral arise from the lower limit of the integral. We can use the heat kernel expansion
\begin{equation}
 \text{Tr} \, e^{-u H}\sim\frac{1}{ u^{d/2}}\left(a_{0}+a_{1/2}u^{1/2}+a_{1}u+\ldots\right)
\end{equation}
to get 
\begin{eqnarray}\label{int}
 e^{-y\overline{\epsilon}_{0}}G(y)&=&\frac{1}{2\sqrt{\pi}}\int_{0}^{\infty}\frac{du}{u^{3/2}}e^{-\frac{y^{2}}{4u}-u}\,\text{Tr}\,e^{-u(m')^{-2}H}\nonumber\\
 &=&\sum_{n=0}^{\infty}2 m'^{d-n}a_{n/2}\left(\frac{y}{2}\right)^{
  \frac{n-d-1}{2}}K_{\frac{-n+d+1}{2}}(y)+\ldots.
\end{eqnarray}
Here, $ K_{\nu}(y) $ is the modified Bessel function of the second kind, and ellipsis stands for regular terms in $y$. 

At this stage, we set $d=3$ and use the series expansion of $ K_{\nu}(y) $ to arrive at the singular value expansion
\begin{equation}\label{intafter}
  e^{-y\overline{\epsilon}_{0}}G(y)\asymp  b_{-4} \,y^{-4}+b_{-3\,}y^{-3}+ O(y^{-2}).
\end{equation}
where the $b$ coefficients are given as
\begin{eqnarray}\label{b}
  b_{-4} &=& \frac{(2m')^{3}}{\sqrt{\pi}}\, \Gamma\left( \frac{4}{2}\right)\, a_{0},\nonumber \\
  b_{-3} &=& \frac{(2m')^{2}}{\sqrt{\pi}} \,\, \Gamma\left( \frac{3}{2}\right)\, a_{1/2},\nonumber  \\
  b_{-2} &=&  \frac{(2m')^{1}}{\sqrt{\pi}}\,\, \Gamma\left( \frac{2}{2}\right) \, \left(-m'^{2}a_{0}+ a_{1}  \right)
\end{eqnarray}
Here we listed only the first few coefficients that we will need in the following. For the Dirichlet problem, the heat kernel coefficients for the operator $-\Delta+U$ in a region $\mathcal{S}$ with boundary $\partial\mathcal{S}$ are given by \cite{Gilkey, Vas}
\begin{eqnarray}\label{a}
  a_{0} &=& \frac{1}{(4\pi)^{3/2}}\int_{\mathcal{S}} dV, \nonumber\\
  a_{1/2} &=& -\frac{1}{4(4\pi)}\int_{\partial \mathcal{S}}dS, \nonumber \\
  a_{1} &=& \frac{1}{6(4\pi)^{3/2}}\left[\int_{\mathcal{S}} dV(-6U+R)+2\int_{\partial \mathcal{S}}dS \,K_{aa}\right],\nonumber\\
  a_{3/2}&=&-\frac{1}{4}\frac{1}{(4\pi)}\frac{1}{96}\int_{\partial \mathcal{S}}dS\,[16(-6U+R)+8R_{aNaN}+7K_{aa}K_{bb}-10K_{ab}K_{ab}].
\end{eqnarray}
In general, coefficients with integer indices contain both bulk and boundary terms (except for $a_{0}$, which is just proportional to the volume of the bulk), and those with half-integer indices consist only of boundary terms.

Here, we use an orthonormal basis $ \{ E_{a},N \} $ adapted to $\partial \mathcal{S}$ where $E_{a}$'s are tangent to $\partial \mathcal{S}$ and $N$ is the inward normal to it. Moreover, $R_{ijkl}$ is the Riemann tensor, $K_{ab}$ is the extrinsic curvature of $\partial \mathcal{S}$ and summation over repeated orthonormal basis indices $a,b$ is assumed.

Using (\ref{intafter}) in (\ref{Gdef}), expanding the exponential factor, and multiplying the series, we get for small $y$;
\begin{equation}
  G(y)\asymp c_{-4}\,y^{-4}+c_{-3}\,y^{-3}+c_{-2}\,y^{-2}+O(y^{-1}),
\end{equation}
where the coefficients of interest to us are
\begin{eqnarray}\label{cb}
  c_{-4}&=& b_{-4},\;\;\;\;c_{-3}=b_{-3}+\overline{\epsilon}_{0}\,b_{-4}\nonumber \\
  c_{-2} &=& b_{-2}+\overline{\epsilon}_{0}\,b_{-3}+\frac{1}{2}\overline{\epsilon}_{0}^{2}\,b_{-4}.
\end{eqnarray}
So
\begin{eqnarray}
  \widetilde{G}(s,-\overline{\mu}_{r})&=&\int_{0}^{\infty}dy\, y^{s-1}e^{\overline{\mu}_{r} y}G(y) \nonumber \\
   &\sim & c_{-4}\frac{\Gamma(s-4)}{(-\overline{\mu}_{r})^{s-4}}+c_{-3}\frac{\Gamma(s-3)}{(-\overline{\mu}_{r})^{s-3}}
   +c_{-2}\frac{\Gamma(s-2)}{(-\overline{\mu}_{r})^{s-2}}+\ldots.
\end{eqnarray}
 Around $-m$ ($m=0,1,2,\ldots$),
\begin{equation}
  \Gamma(s-n)\sim \frac{(-1)^{m}}{m!}\frac{1}{s-n+m},
\end{equation}
and we also have the simple pole of $\zeta$ at $s=1$
\begin{equation}\label{zeta}
\zeta(s)\sim \frac{1}{s-1}+\gamma,
\end{equation}
where $\gamma$ is the Euler-Mascheroni constant. Thus, the poles of $\zeta(s)\widetilde{G}(s,-\overline{\mu}_{r})$ are seen to be the integers $3,2,1,\ldots$. Here, all the poles are simple, except the one at $s=1$, which is a double pole. So we have the following residues:\\
Around $s=4$:
\begin{equation}
   \zeta(s)\widetilde{G}(s,-\overline{\mu}_{r})\asymp \zeta(4)c_{-4}\frac{1}{s-4},
\end{equation}
around $s=3$:
\begin{equation}
   \zeta(s)\widetilde{G}(s,-\overline{\mu}_{r})\asymp \zeta(3)(c_{-4}\overline{\mu}_{r}+c_{-3})\frac{1}{s-3},
\end{equation}
around $s=2$:
\begin{equation}
   \zeta(s)\widetilde{G}(s,-\overline{\mu}_{r})\asymp \zeta(2)(\frac{1}{2}c_{-4}\overline{\mu}_{r}^{2}+c_{-3}\overline{\mu}_{r}+c_{-2})\frac{1}{s-2}.
\end{equation}
and around $s=1$ we have
\begin{eqnarray}
  \zeta(s)\widetilde{G}(s,-\overline{\mu}_{r})\asymp \left(\frac{1}{2}c_{-3}\overline{\mu}_{r}^{2}+c_{-2}\overline{\mu}_{r}+c_{-1}\right)\frac{1}{(s-1)^{2}}\nonumber\\
  +\left[FP\widetilde{G}(s=1,-\overline{\mu}_{r})+\gamma\left(\frac{1}{2}c_{-3}\overline{\mu}_{r}^{2}+c_{-2}\overline{\mu}_{r}+c_{-1}\right)\right]\frac{1}{s-1}.
\end{eqnarray}
Here, $FP$ means the finite part of the integral. Since at $s=1$ we have a double pole, there will be, according to the general formula (\ref{mellinasy}), a logarithmic term in the small $y$ (high $T$) expansion of $\mathcal{F}$.

Converting the coefficients via (\ref{mur}), (\ref{b}) and (\ref{cb}), and using (\ref{mellinasy}), we arrive at the high $T$ expansion
\begin{eqnarray}\label{free0}
\mathcal{F} &=& -\frac{\zeta(4)}{\sqrt{\pi}} \, 8 \, a_{0}\, T^{4}  -\frac{\zeta(3)}{\sqrt{\pi}} \left[ 8\mu \,a_{0}+2\,\sqrt{\pi}\,a_{1/2}\right] \, T^{3}+ \nonumber\\
&&- \frac{\zeta(2)}{\sqrt{\pi}}\left[ (4\mu^{2}-2m'^{2})a_{0}+2\sqrt{\pi}\mu  a_{1/2}+2 a_{1}\right] \, T^{2}+\nonumber\\
&&+\frac{1}{2\sqrt{\pi}} \left[ \left(\frac{8}{3} \mu^{3}-4 m'^{2}\mu \right)  a_{0}+ 2\sqrt{\pi} (\mu^{2}-m'^{2})a_{1/2} +4\mu a_{1} +2 \sqrt{\pi}\, a_{3/2} \right] T\log \left(\frac{T}{m'}\right)+ \nonumber\\
 &&- \frac{1}{2\sqrt{\pi}}  \Biggl\{ \gamma \left[ \left(\frac{8}{3} \mu^{3}-4 m'^{2}\mu \right)  a_{0}+ 2\sqrt{\pi} (\mu^{2}-m'^{2})a_{1/2} +4\mu a_{1} +2\sqrt{\pi}\, a_{3/2} \right] +  \nonumber\\
  &&+ FP\widetilde{G}\left(s=1,\frac{\epsilon_0-\mu}{m'}\right) \Biggr\} T+O(T^{0}).
\end{eqnarray}

From $ \mathcal{F}_{cs}(\beta,\mu)=\mathcal{F}_{rs}(\beta,\mu)+\mathcal{F}_{rs}(\beta,-\mu)$ we then have, for the charged scalar field,
\begin{eqnarray}\label{cbosef}
  \mathcal{F}_{cs}&=&   -\frac{\zeta(4)}{\sqrt{\pi}} \, 16 \, a_{0}\, T^{4} - \frac{\zeta(3)}{\sqrt{\pi}} 4 a_{1/2} \, T^{3}+ \nonumber\\
&&- \frac{\zeta(2)}{\sqrt{\pi}}4\left[ (2\mu^{2}-m^{'2})a_{0}+ a_{1}\right] \, T^{2}+2\left[  (\mu^{2} -m^{'2}) a_{1/2}+ a_{3/2}\right]   T \log \left(\frac{T}{m'}\right)+\nonumber\\
&&-  \left\{ \gamma \left[  2(\mu^{2}-m^{'2}) a_{1/2}+2 \, a_{3/2} \right]   +\frac{1}{\sqrt{\pi}}FP\widetilde{G}\left(1,\frac{\epsilon_0-\mu}{m'}\right) \right\} T +O(T^{0}).
\end{eqnarray}

\subsection{Free Dirac Field}
A very similar analysis can be done for fermions by using the Dirac equation instead of the Klein-Gordon equation. For the Dirac field, we have
\begin{equation}
  \mathcal{F}_{Dirac}= -\frac{2}{\beta}\sum_{\sigma}
  \left[\log(1+e^{-\beta(\epsilon_{\sigma}-\mu)})+\log(1+e^{-\beta(\epsilon_{\sigma}+\mu)})\right].
\end{equation}
Again, expanding the logarithms we get
\begin{eqnarray}
  \mathcal{F}_{Dirac}(\mu) &=& -4\sum_{\sigma}\sum_{k=1}^{\infty}\frac{(-1)^{k+1}}{k\beta}\cosh (k\beta\mu) e^{-k\beta\epsilon_{\sigma}}.
  \end{eqnarray}
The rest of the analysis parallels the bosonic case. Due to the the alternating nature of the series expansion, the Mellin transform now generates the Dirichlet $\eta$ function instead of the Riemann $\zeta$.
\begin{eqnarray}\label{fermif}
  \mathcal{F}_{Dirac}&=&  -2\frac{\eta(4)}{\sqrt{\pi}} \, 16 \, a_{0}\, T^{4} -2 \frac{\eta(3)}{\sqrt{\pi}} 4 a_{1/2} \, T^{3}+ \nonumber\\
&&-2 \frac{\eta(2)}{\sqrt{\pi}}4\left[ (2\mu^{2}-m^{'2})a_{0}+ a_{1}\right] \, T^{2}+ \nonumber\\
&&-2\eta(1)\left[ 2 (\mu^{2} -m^{'2}) a_{1/2}+ 2\,  a_{3/2}\right] \, T +O(T^{0}),
\end{eqnarray}
where
\begin{equation}
  \eta(s)\equiv\sum_{k=1}^{\infty}\frac{(-1)^{k-1}}{k^{s}}.
\end{equation}
Unlike the bosonic case we do not have a logarithmic term in this expansion since the $\eta$ function is entire and consequently the poles of the Mellin transform are all simple.

\section{Boundary and Geometric Effects in Black Hole Backgrounds}

 In what follows, we will focus on the case of vanishing chemical potential. Setting $\mu=0$ and using the explicit forms of the heat kernel coefficients (\ref{a}) in (\ref{free0}) we get the free energy of the scalar field as
\begin{eqnarray}\label{free}
  \mathcal{F}_{rs} &=&  - \zeta(4)\frac{1}{\pi^{2}}\,V T^{4}+\zeta(3)\frac{1}{8\pi}\,A T^{3}+\nonumber\\
  &&-\frac{\zeta(2)}{4\pi^{2}}\left\{\int_{\mathcal{S}}dV\,(-F)\left[m^{2}+\left(\xi-\frac{1}{6}\right)\mathcal{R}\right]+\int_{\partial \mathcal{S}}dS\,
   \frac{K_{aa}}{3} \right\}  T^{2}\nonumber\\
  && -\frac{1}{16\pi}\int_{\partial \mathcal{S}}dS\,\left\{-F\left[m^{2}+\left(\xi-\frac{1}{6}\right)\mathcal{R}\right]\right.\nonumber\\
  &&\left.+\frac{1}{96}[8R_{NaNa}+7K_{aa}K_{bb}-10K_{ab}K_{ab}]\right\}T\log\frac{T}{m'}+O(T).
\end{eqnarray}
Consequently, the entropy is given as
\begin{eqnarray}\label{sexp}
  S_{rs} &=&-\frac{\partial\mathcal{F}_{rs}}{\partial T}=\frac{4\zeta(4) }{\pi^{2}}\,V T^{3}-\frac{3\zeta(3)}{8\pi}\,AT^{2}\nonumber\\&&+ \frac{\zeta(2)}{2\pi^{2}}\left\{\int_{\mathcal{S}}dV\,(-F)\left[m^{2}+\left(\xi-\frac{1}{6}\right)\mathcal{R}\right]+\int_{\partial \mathcal{S}}dS\,
   \frac{K_{aa}}{3} \right\}T\nonumber\\
   && +\frac{1}{16\pi}\int_{\partial \mathcal{S}}dS\,\left\{-F\left[m^{2}+\left(\xi-\frac{1}{6}\right)\mathcal{R}\right]+\frac{1}{96}[8R_{NaNa}+7K_{aa}K_{bb}-10K_{ab}K_{ab}]\right\}\log\frac{T}{m'}\nonumber\\
   &&+O(T^{0}).
\end{eqnarray}
Note the cancelation of the factors of $m'$ except in the logarithmic term; we will come back to this point in Sec.3.2. From (\ref{fss}) we have for the charged scalar field,
\begin{equation}
  \mathcal{F}_{cs}=2\mathcal{F}_{rs}\;\;\;\textrm{and}\;\;\;S_{cs}=2S_{rs}.
\end{equation}
On the other hand, for the Dirac field with $\mu=0$, using (\ref{a}) in (\ref{fermif}), we obtain
\begin{eqnarray}\label{freedirac}
  \mathcal{F}_{Dirac} &=&  - \frac{4\eta(4)}{\pi^{2}}\,V T^{4}+\frac{4\eta(3)}{8\pi}\,AT^{3}+\nonumber\\
  &&-\frac{4\eta(2)}{4\pi^{2}}\left\{\int_{\mathcal{S}}dV\,(-F)\left[m^{2}+\left(\xi-\frac{1}{6}\right)\mathcal{R}\right]+\int_{\partial \mathcal{S}}dS\,
   \frac{K_{aa}}{3} \right\}  T^{2}\nonumber\\
  && -\frac{4\eta(1)}{16\pi}\int_{\partial \mathcal{S}}dS\,\left\{-F\left[m^{2}+\left(\xi-\frac{1}{6}\right)\mathcal{R}\right]\right.\nonumber\\
  &&\left.+\frac{1}{96}[8R_{NaNa}+7K_{aa}K_{bb}-10K_{ab}K_{ab}]\right\}T+O(T^{0})
\end{eqnarray}
and
\begin{eqnarray}\label{sexpdirac}
  S_{Dirac}&=&\frac{16\eta(4) }{\pi^{2}}\,V T^{3}-\frac{12\eta(3)}{8\pi}\,AT^{2}\nonumber\\ &+& \frac{4\eta(2)}{2\pi^{2}}\left\{\int_{\mathcal{S}}dV\,(-F)\left[m^{2}+\left(\xi-\frac{1}{6}\right)\mathcal{R}\right]+\int_{\partial \mathcal{S}}dS\,
   \frac{K_{aa}}{3} \right\}T\nonumber\\
   &+&\frac{4\eta(1)}{16\pi}\int_{\partial \mathcal{S}}dS\,\left\{-F\left[m^{2}+\left(\xi-\frac{1}{6}\right)\mathcal{R}\right]+\frac{1}{96}[8R_{NaNa}+7K_{aa}K_{bb}-10K_{ab}K_{ab}]\right\}\nonumber\\
   &+&O(T^{-1}).
\end{eqnarray}

In this section, we consider Schwarzschild, RN, extreme RN, and dilatonic metrics. All these metrics are of the form
\begin{equation}
  ds^{2}=-F(r)dt^{2}+F^{-1}(r)dr^{2}+r(r-a)[d\theta^{2}+\sin^2\theta\, d\phi^{2}],\;\;\;\;\;r> a
\end{equation}
where $a=0$ for Schwarzschild and RN metrics. We will take $\mathcal{S}$ as the "spherical" shell defined by $r_{1}\leq r\leq r_{2}$ with $r_{1}=r_{H}+\epsilon$, where $r_{H}$ is the radial coordinate of the horizon, $\epsilon$ is the brick wall cut-off and $r_{2}$ is an infrared cut-off.

The optical metric is
\begin{equation}
  d\overline{s}^{2}=F^{-2}(r)dr^{2}+r(r-a)F^{-1}(r)\left(d\theta^{2}+\sin^2\theta\, d\phi^{2}\right).
\end{equation}
The quantities needed for the calculation of the heat kernel coefficients (\ref{a}) are then as follows.

The volume element is given as
 \begin{equation}\label{volume}
  dV=F^{-2}r(r-a)\sin\theta \, dr\wedge d\theta\wedge d\phi.
\end{equation}
The area element of a constant $r$ surface is given as
\begin{equation}\label{area}
  dS_{r}= F^{-1}r(r-a)\sin\theta \,d\theta\wedge d\phi,  
\end{equation}
The inward normal to the brick wall is given by the value at $r=r_{1}$ of the vector field
\begin{equation}
  N = F\partial_{r},
\end{equation}
and the trace of the corresponding extrinsic curvature is
\begin{equation}\label{extrinsic}
   K_{aa} =-\nabla\cdot N=F'(r)-\frac{(2r-a)F(r)}{r(r-a)}.
\end{equation}
Consider the orthonormal basis
\begin{eqnarray}\label{normal}
  N_{(r)} = F\partial_{r},\;\;E_{1} = \frac{\sqrt{F}}{\sqrt{r(r-a)}}\partial_{\theta},\;\;E_{2} = \frac{\sqrt{F}}{\sqrt{r(r-a)}\sin\theta}\partial_{\phi}.
\end{eqnarray}
This basis is adapted to constant $r$ surfaces, that is, $N$ is normal and $E$'s are tangential to constant $r$ surfaces. On the brick wall at $r=r_{1}$, $N$ is the inward looking normal.
The extrinsic curvature of the inner wall at $r=r_{1}$ is given by
\begin{eqnarray}\label{extrinsicmat}
K_{ab}=N\cdot(\nabla_{E_{a}}E_{b})=N_{\nu}E_{a}^{\mu}\partial_{\mu}E_{b}^{\nu}+\Gamma_{\mu\sigma}^{\nu}E_{a}^{\mu}E_{b}^{\sigma}N_{\nu}=\Gamma_{\mu\sigma}^{\nu}E_{a}^{\mu}E_{b}^{\sigma}N_{\nu}.
\end{eqnarray}
More explicitly, using $N_{\nu}=F^{-1}\delta_{\nu}^{0}$,
\begin{eqnarray}\label{kas}
  K_{11}=\frac{\Gamma^{r}_{\theta\theta}}{r(r-a)},\;\;\; K_{22}=\frac{\Gamma^{r}_{\phi\phi}}{r(r-a)\sin^{2}\theta},\;\;\; K_{12}=K_{21}= \frac{\Gamma^{r}_{\theta\phi}}{r(r-a)\sin\theta}.
\end{eqnarray}
We also have
\begin{eqnarray}\label{ranan}
  R_{aNaN} &=& \frac{F^{3}}{r(r-a)} R_{\theta r \theta r}+ \frac{F^{3}}{r(r-a)\sin^{2}\theta}  R_{\phi r \phi r}.
\end{eqnarray}
Moreover, the proper length cut-off is given by
\begin{eqnarray}\label{deltaepsilon}
 \delta&=&\int_{r_{H}}^{r_{H}+\epsilon}\frac{dr}{\sqrt{F(r)}}
 \end{eqnarray}
 Expanding $F(r)$ around $r_{H}$ we get,
 \begin{eqnarray}\label{deleps}
 \delta&=&\frac{1}{\sqrt{F'(r_{H})}}\int_{r_{H}}^{r_{H}+\epsilon}\frac{dr}{\sqrt{r-r_{H}}\left[1+\frac{F''(r_{H})}{2F'(r_{H})}(r-r_{H})+O((r-r_{H})^{2})\right]^{1/2}}\nonumber\\
 &=& \sqrt{\frac{2\epsilon}{\kappa}}\left[1+O(\epsilon)\right],
\end{eqnarray}
where 
\begin{equation}
    \kappa=\frac{F'(r_{H})}{2}
\end{equation}
is the surface gravity of the horizon. Here we assume $\kappa\neq 0$, which holds for all non-extremal metrics considered. Squaring both sides and solving for $\epsilon$ we get
\begin{equation}
  \epsilon=\frac{\kappa\delta^2}{2}+O\left(\left(\frac{\kappa\delta^2}{2}\right)^2\right).
\end{equation}
Also note that
\begin{equation}\label{oneover}
\frac{1}{\epsilon}=\frac{2}{\kappa\delta^2}+O\left(1\right),\;\;\;\;\;\log\epsilon=\log\frac{\kappa\delta^2}{2}+
O\left(\frac{\kappa\delta^2}{2}\right).
\end{equation}

\subsection{Entropy of the scalar field}
In the Appendix we compute the expression (\ref{sexp}) for the entropy $S_{rs}$ of the real scalar field at the Hawking temperature $T_{H}$ for Schwarzschild, RN and dilaton black holes. 

We find that in all these cases, $S_{rs}$ is given by an expression of the form
\begin{equation}\label{sexprs}
  S_{rs}=B_{rs}A_{H}\frac{1}{\delta^{2}}+C_{rs}\log\frac{\kappa\delta^{2}}{2(r_2-r_H)}.
\end{equation}
Here, $A_{H}$ is the horizon area, $\delta$ is the proper radial distance of the brick wall from the horizon, and $\kappa$ is the surface gravity of the horizon. 

The coefficient $B_{rs}$ is the same for all cases and is given as
\begin{equation}\label{BRealScal}
  B_{rs}=\frac{1}{4\pi^{3}}\left[\frac{\zeta(4)}{\pi^{2}}-\frac{3}{8}\zeta(3)+\frac{2}{3}\zeta(2)-\frac{\pi^{2}}{48}\, \log \frac{\kappa}{2\pi m'}\right].
\end{equation}

As for the sub-leading contribution, the coefficient $C$ for the Schwarzschild metric is 
\begin{equation}
  C_{rs}^{S}=\frac{1}{\pi^{2}}\left[ -\frac{\zeta(4)}{\pi^{2}}+\frac{\zeta(2)}{8\pi}m^{2}\,A_{H}\right],
\end{equation}
for the RN metric it is given by
\begin{equation}\label{CRNREALSCAL}
  C_{rs}^{RN}=\frac{1}{\pi^{2}}\left[-\frac{\zeta(4)}{\pi^{2}} \, \frac{2r_{+}-3r_{-}}{2r_{+}}+\frac{\zeta(2)}{8\pi}m^{2}\, A_{H} \right] ,
\end{equation}
and for the dilatonic metric we have,
\begin{equation}
  C_{rs}^{D}=\frac{1}{\pi^{2}}\left[ -\frac{\zeta(4)}{\pi^{2}}\frac{(8M-3a)}{8M}+\frac{\zeta(2)}{8\pi}m^{2}\,A_{H}\right].
\end{equation}
The terms proportional to $\zeta(4)$ (in both $B$ and $C$ coefficients) originate from the leading $O(T^{3})$ term of (\ref{sexp}) which is proportional to the optical volume and are in complete agreement with the results of \cite{Solodukhin} and \cite{Alwis}.

The result for the complex scalar field can easily be derived from (\ref{fss}),
\begin{equation}
  S_{cs}=2S_{rs}=B_{cs}A_{H}\frac{1}{\delta^{2}}+C_{cs}\log\frac{\kappa\delta^{2}}{2(r_2-r_H)},
\end{equation}
where $B_{cs}=2B_{rs}$ and $C_{cs}=2C_{rs}$.

As shown in the Appendix, the leading horizon divergence receives contributions from both the bulk and boundary terms of the heat kernel coefficients. On the other hand, a general boundary term in a heat kernel coefficient involves a surface integral over the brick wall, whose equation is $r=r_{H}+\epsilon$, of a polynomial function of curvature invariants (both intrinsic and extrinsic) and potential $U_{1}$, which in turn depend only on the radial coordinate $r$ and do not have any logarithmic behavior. Thus, we conclude that the boundary terms of the heat kernel coefficients contribute only to the leading horizon divergence. In other words, the sub-leading horizon divergence is independent of the boundary condition and is determined solely by the bulk contributions to the heat kernel expansion.

\subsection{Leading Horizon Divergence}

 As noted in the Introduction, the leading order horizon divergence in $S$ given in  (\ref{sexprs}) leads to the renormalization of the gravitational constant \cite{Davies, Susskind}.
On the other hand, the coefficient $B_{rs}$ given in (\ref{BRealScal}) is a numerical factor, except for the logarithmic term in $\kappa$. Going back to (\ref{mp}), the very first place we introduced the arbitrary parameter $m'$, we note that the resulting expansion should be independent of $m'$. In fact, our results for the coefficients $B$ and $C$ explicitly show this independence, except for the logarithmic term $\log(\kappa/m')$. To better understand this term, consider one more order in the high $T$ expansion of $S$. There will be no need for explicit forms of the heat coefficients, so we simply take the negative $T$ derivative of (\ref{free0}) and get 
\begin{eqnarray}\label{ent0}
S &=& \frac{\zeta(4)}{\sqrt{\pi}} \, 32 \, a_{0}\, T^{3}  +\frac{3\zeta(3)}{\sqrt{\pi}} \left[ 8\mu \,a_{0}+2\,\sqrt{\pi}\,a_{1/2}\right] \, T^{2} \nonumber\\
&&+ \frac{2\zeta(2)}{\sqrt{\pi}}\left[ (4\mu^{2}-2m'^{2})a_{0}+2\sqrt{\pi}\mu  a_{1/2}+2 a_{1}\right] \, T\nonumber\\
&&-\frac{1}{2\sqrt{\pi}} \left[ \left(\frac{8}{3} \mu^{3}-4 m'^{2}\mu \right)  a_{0}+ 2\sqrt{\pi} (\mu^{2}-m'^{2})a_{1/2} +4\mu a_{1} +2 \sqrt{\pi}\, a_{3/2} \right] \log \left(\frac{T}{m'}\right)\nonumber\\
 &&+ \frac{1}{2\sqrt{\pi}}  \Biggl\{ (\gamma-1) \left[ \left(\frac{8}{3} \mu^{3}-4 m'^{2}\mu \right)  a_{0}+ 2\sqrt{\pi} (\mu^{2}-m'^{2})a_{1/2} +4\mu a_{1} +2\sqrt{\pi}\, a_{3/2} \right]   \nonumber\\
  &&+ FP\widetilde{G}\left(1,\frac{\epsilon_{0}-\mu}{m'}\right) \Biggr\} +O(T^{-1}).
\end{eqnarray}
 Writing  $ \log (T/m')=\log T-\log m'$, we see that $\log m'$ term must be canceled by a contribution coming from the $O(T)$ term of the expansion. But then, on dimensional grounds, the same $O(T)$ term must also provide a logarithmic term in an intrinsic dimensionful parameter which can then be combined with the remaining $\log T$ term to produce the logarithm of a dimensionless quantity. In other words, there must be an intrinsic $m'$ that, on dimensional grounds, is given as a combination of the parameters of the background metric and possibly the mass of the quantum field. If it turns out that this intrinsic value of $m'$ is a numerical multiple of $T_{H}$ (or of $\kappa$), then the logarithmic term in the coefficient $B$ would become a numerical factor when we evaluate the entropy at $T_{H}$, and consequently $B$ itself would become a pure number. In Sec.4 we will show that this is indeed the case in the near-horizon geometry. Investigation of the situation in exact backgrounds, which requires a detailed analysis of the term $FP\widetilde{G}(1,(\epsilon_{0}-\mu)/m')$, will be carried out elsewhere. 

\subsection{Sub-Leading Horizon Divergence}
Consider the expression (\ref{CRNREALSCAL}) for $C^{RN}_{rs}$. For RN metric we have $\kappa/2\pi=(r_{+}-r_{-})/(4\pi r_{+}^{2})$ and $A_{H}=4\pi r_{+}^{2} $. So, $r_{+}-r_{-}=\kappa A_{H}$ and $r_{+}=\sqrt{A_{H}}/(2\sqrt{\pi})$. Using these, we can express $C^{RN}_{rs}$ in terms of $\kappa$ and $A_{H}$ as
\begin{equation}
  C_{rs}^{RN}=C_{rs}(m,\kappa,A_{H})
\end{equation}
where we defined the function
\begin{equation}\label{CKA}
  C_{rs}(m,\kappa,A_{H})=-\frac{\zeta(4)}{\pi^{4}}\left[\frac{3}{2\sqrt{\pi}}\kappa \sqrt{A_{H}}-\frac{1}{2}\right]+\frac{\zeta(2)}{8\pi^{3}}m^{2}A_{H}.
\end{equation}
Evaluating this function for the Schwarzschild metric, for which $\kappa=1/(4M)$ and $A_{H}=16\pi M^{2}$, we get the correct $C_{rs}^{S}$ coefficient
\begin{equation}
  C_{rs}\left(m,\frac{1}{4M},16\pi M^{2}\right)=-\frac{\zeta(4)}{\pi^{4}}+\frac{\zeta(2)}{8\pi^{3}}m^{2}A_{H}=C_{rs}^{S},
\end{equation}
For the extreme RN black hole, for which $\kappa=0$ and $A_{H}=4\pi r_{+}^{2}$, we also find the correct coefficient,
\begin{equation}
  C_{rs}\left(m,0,4\pi r_{+}^{2}\right)=\frac{\zeta(4)}{2\pi^{4}}+\frac{\zeta(2)}{8\pi^{3}}m^{2}A_{H}=C_{rs}^{ERN},
\end{equation}
For the dilaton black hole, we have a slight complication due to the dilaton charge $a$. In this case, we have $\kappa=1/(4M)$, $A_{H}=4\pi\, 2M(2M-a)$ and using these we can write
\begin{eqnarray}
  C_{rs}^{D}&=&\frac{\zeta(4)}{\pi^{4}}\left[-\frac{8M-3a}{8M}\right]+\frac{\zeta(2)}{8\pi^{3}}m^{2}A_{H}\nonumber\\
  &=&\frac{\zeta(4)}{\pi^{4}}\left[\frac{1}{2}-\frac{3}{2\sqrt{\pi}}\kappa \sqrt{A_{H}}\frac{1-a\kappa}{\sqrt{1-2a\kappa}}\right]+\frac{\zeta(2)}{8\pi^{3}}m^{2}A_{H}.
\end{eqnarray}\\
So, we arrive at a general expression that holds for all the metrics considered,
\begin{equation}\label{CRSMKAA}
  C_{rs}(m,\kappa,A_{H},a)=\frac{\zeta(4)}{\pi^{4}}\left[\frac{1}{2}-\frac{3}{2\sqrt{\pi}}\kappa \sqrt{A_{H}}\frac{1-a\kappa}{\sqrt{1-2a\kappa}}\right]+\frac{\zeta(2)}{8\pi^{3}}m^{2}A_{H}.
\end{equation}
Note that for $a=0$,  $C_{rs}(m,\kappa,A_{H},0)=C_{rs}(m,\kappa,A_{H})$.

Thus, in all the cases we consider the sub-leading horizon divergence in entropy is given by the same function of $m$, $A_{H}$, $\kappa$, and $a$
\begin{equation}
  S_{rs}=B_{rs}A_{H}\frac{1}{\delta^{2}}+C_{rs}(m,\kappa,A_{H},a)\log\frac{\kappa\delta^{2}}{2(r_2-r_H)}.
\end{equation}

\subsection{Entropy in the Extreme Reissner-Nordstr\"{o}m Geometry}

Thermodynamics in the extreme RN background is not well defined, since, as shown in \cite{Alwis}, to leading order in the high-temperature expansion the entropy involves, on top of the usual $O(\epsilon^{-1})$ and $O(\log\epsilon)$ terms, also $O(\epsilon^{-2})$ and $O(\epsilon^{-3})$ divergences that render it non-renormalizable. Here we will examine the entropy in the extreme background following the proposal of references \cite{Alwis0} and \cite{Alwis}, according to which the thermodynamics in the extreme RN case should be regarded as the limiting case of the thermodynamics in the non-extreme background. More precisely, the physically relevant regime is $r_{+}\gtrsim r_{-}$, since even a single quantum exchange between the black hole and the field spoils the extremality of the former, and the approximation $r_{-}\approx r_{+}$ is to be made after the small $\epsilon$ asymptotics is derived.

In the RN geometry
$\kappa=(r_{+}-r_{-})/2r_{+}^{2}$ and therefore according to (\ref{deleps}) $\delta$ depends of $r_{+}-r_{-}$. So, in order to make the factors of $r_{+}-r_{-}$ manifest, we express $S$ in terms of  $\epsilon$ instead of the proper length cut-off $\delta$. As shown in Appendix A.3 this then gives  \begin{eqnarray}
  S_{rs}&=&\frac{1}{4\pi^{3}}\left[\frac{\zeta(4)}{\pi^{2}}-\frac{3}{8}\zeta(3)+\frac{2}{3}\zeta(2)-\frac{\pi^{2}}{48}\, \log \frac{r_{+}-r_{-}}{A_{H} m'}\right]\frac{(r_{+}-r_{-})A_{H}}{4 r_{+}^{2}}\frac{1}{\epsilon}+\nonumber\\&& +  \frac{1}{\pi^{2}}\left[-\frac{\zeta(4)}{\pi^{2}} \, \frac{2(r_{+}-r_{-})-r_{-}}{2r_{+}}+\frac{\zeta(2)}{8\pi}m^{2}\, A_{H} \right]\log\frac{\epsilon}{(r_2-r_H)},
\end{eqnarray} 
We note that $\zeta(4)$ term of the leading horizon divergence, which arises from the leading term of the high-temperature expansion, agrees with the WKB result of \cite{Qiu} evaluated at $T_{H}$.   

In fact, at this point we can take the actual $r_{-}\rightarrow r_{+}$ limit. We find that $O(\epsilon^{-1})$ horizon divergence vanishes and the logarithmic divergence becomes the leading term,
\begin{equation}
  S_{rs}=\frac{1}{\pi^{2}}\left[\frac{\zeta(4)}{2\pi^{2}} +\frac{\zeta(2)}{8\pi}m^{2}\, A_{H}\right]\log\frac{\epsilon}{(r_2-r_H)},
\end{equation}
So, Newton's constant does not get renormalized, while the logarithmic divergence is expected to be renormalized by the counterterms in the gravitational action.

\subsection{Entropy of the Dirac field}
Comparing $(\ref{sexp})$ and $(\ref{sexpdirac})$ we see that $S_{Dirac}$ can be obtained from $S_{rs}$ by the substitutions $\zeta(k)\rightarrow 4\eta(k)$ and $\ln \kappa/2\pi m'\rightarrow -\eta(1)$. Thus,
\begin{equation}
  S_{Dirac}=B_{Dirac}A_{H}\frac{1}{\delta^{2}}+C_{Dirac}(m,\kappa,A_{H},a)\log\frac{\kappa\delta^{2}}{2(r_2-r_H)},
\end{equation}
where
\begin{eqnarray}
  B_{Dirac} &=&\frac{1}{\pi^{3}}\left[\frac{\eta(4)}{\pi^{2}}-\frac{3}{8}\eta(3)+\frac{2}{3}\eta(2)+\frac{\pi^{2}}{48}\,\eta(1)\right]  \\
  C_{Dirac}(m,\kappa,A_{H},a) &=&  \frac{4\eta(4)}{\pi^{4}}\left[\frac{1}{2}-\frac{3}{2\sqrt{\pi}}\kappa \sqrt{A_{H}}\frac{1-a\kappa}{\sqrt{1-2a\kappa}}\right]+\frac{\eta(2)}{2\pi^{3}}m^{2}A_{H}.
\end{eqnarray}
Computing $B_{Dirac}$ explicitly, we get  
 \begin{eqnarray}
   B_{Dirac} \approx 1.447\times 10^{-2}.
\end{eqnarray}

\section{Entropy in the Near-Horizon Geometry}

In this section, we will consider only the real scalar field. If needed, our results can easily be generalized to the other cases. 

Near the horizon, the metric 
\begin{equation}
    ds^{2}=-F(r)dt^{2}+F^{-1}(r)dr^{2}+r(r-a)(d\theta^{2}+\sin^{2}\theta d\phi^{2})
\end{equation}
is approximated as
\begin{equation}
    ds_{NH}^{2}=-2\kappa\rho' dt^{2}+\left(2\kappa \rho'\right)^{-1} d\rho'^{2}+r_{H}(r_{H}-a)(d\theta^{2}+\sin^{2}\theta d\phi^{2}).
\end{equation}
Here, $\rho'=r-r_{H}$, $r_{H}$ is the radial coordinate of the event horizon and $\kappa=F'(r_{H})/2$ is the surface gravity. The brick wall is located at $\rho'=\epsilon$. The proper radial distance of the brick wall to the horizon is given by
\begin{equation}
  \delta=\int_{0}^{\epsilon}\frac{d\rho'}{\sqrt{{2\kappa \rho'}}}=\sqrt{\frac{2\epsilon}{\kappa}}.
\end{equation}

The scalar curvature of the near-horizon metric is
\begin{equation}
    \mathcal{R}=\frac{2}{r_{H}(r_{H}-a)}=\frac{8\pi}{A_{H}}.
\end{equation}
Here, $A_{H}=4\pi r_{H}(r_{H}-a)$ is the area of the event horizon.

The corresponding optical metric is
\begin{equation}
    d\overline{s}_{NH}^{2}=\frac{1}{4\kappa^{2}}\left[\frac{d\rho'^{2}}{\rho'^{2}}+\frac{2\kappa r_{H}(r_{H}-a)}{\rho'}(d\theta^{2}+\sin^{2}\theta d\phi^{2})\right].
\end{equation}
Upon the change of variable
\begin{equation}\label{ro}
    \rho=\sqrt{\frac{2\pi\rho'}{\kappa A_{H}}},
\end{equation}
we get
\begin{equation}
    d\overline{s}_{NH}^{2}=\frac{1}{\kappa^{2}}d\tilde{s}^{2},
\end{equation}
where
\begin{equation}\label{metilde}
    d\tilde{s}^{2}=\frac{d\rho^{2}}{\rho^{2}}+\frac{1}{4\rho^{2}}(d\theta^{2}+\sin^{2}\theta d\phi^{2}).
\end{equation}
 Note that in these coordinates the brick wall is situated at
\begin{equation}\label{pos}
    \rho(\epsilon)=\sqrt{\frac{2\pi\epsilon}{\kappa A_{H}}}=\sqrt{\frac{\pi}{A_{H}}}\delta.
\end{equation}
Also note that the scalar curvature of the optical metric $d\overline{s}_{NH}^{2}$ is
\begin{equation}
R=\kappa^{2}\widetilde{R},
\end{equation}
where $\widetilde{R}$ is the scalar curvature of $d\tilde{s}^{2}$ which is given explicitly by
\begin{equation}
   \widetilde{R}=-6+8\rho^{2}.
 \end{equation}
From now on, we will denote geometric quantities derived from the metric $d\widetilde{s}^{2}$ with a tilde. Similarly, the Laplacian of the optical metric is related to the Laplacian of $d\tilde{s}^{2}$ as
\begin{equation}
    \Delta=\kappa^{2}\widetilde{\Delta}.
\end{equation}
Proceeding as in Sec.2 we get
\begin{eqnarray}
  U_{1}&=&\frac{1}{6}R+2\kappa\rho'\left[m^{2}+\left(\xi-\frac{1}{6}\right)\mathcal{R}\right]\nonumber\\
  &=& \kappa^{2}\widetilde{U}_{1},
\end{eqnarray}
where, in terms of dimensionless coordinate $\rho$, we have 
\begin{equation}\label{pot1}
    \widetilde{U}_{1}=\left(\frac{4\rho^{2}}{3}-1 \right)+\frac{A_{H}}{\pi}\rho^{2}\left[m^{2}+\left(\xi-\frac{1}{6}\right)\frac{8\pi}{A_{H}}\right].
\end{equation}
Thus, $H_{1}$ scales like $\kappa^{2}$,
\begin{equation}
    H_{1}=\kappa^{2}\widetilde{H}_{1},\qquad  \widetilde{H}_{1}=-\widetilde{\Delta}+\widetilde{U}_{1}.
\end{equation}
Defining
\begin{equation}\label{pot2}
  \widetilde{U}=\widetilde{U}_{1}-1,\qquad \widetilde{H}=-\widetilde{\Delta}+\widetilde{U}, 
\end{equation}
we get
\begin{equation}
  \widetilde{H}_{1}=\widetilde{H}+1,\qquad H_{1}=\kappa^{2}\widetilde{H}+ \kappa^{2},
\end{equation}
and
\begin{eqnarray}
  \sum_{\sigma}e^{-k\beta\epsilon_{\sigma}} = \frac{k\beta }{2\sqrt{\pi}}\int_{0}^{\infty}\frac{dv}{v^{3/2}}\,e^{-\frac{(k\beta )^{2}}{4v}}e^{-v\kappa^{2} }\,\text{Tr}\, e^{-v\kappa^{2}\widetilde{H}}.
\end{eqnarray}
Upon the change of variable $u=v\kappa^{2}$ we arrive at
\begin{equation}
 \sum_{\sigma}e^{-k\beta \epsilon_{\sigma}}=\frac{k\beta \kappa }{2\sqrt{\pi}}\int_{0}^{\infty}\frac{du}{u^{3/2}}\,e^{-\frac{(k\beta \kappa )^{2}}{4u}}e^{-u}\,\text{Tr}\, e^{-u\widetilde{H}}.
\end{equation}
So, the free energy is
\begin{eqnarray}\label{FreeNearH}
 \mathcal{F} =  -\sum_{k=1}^{\infty}\frac{1}{k\beta }\sum_{\sigma}e^{-k\beta \epsilon_{\sigma}}= -\frac{\kappa }{2\sqrt{\pi}}\sum_{k=1}^{\infty}\int_{0}^{\infty}\frac{du}{u^{3/2}}\,e^{-\frac{(k\beta \kappa)^{2}}{4u}}e^{-u}\,\text{Tr}\, e^{-u\widetilde{H}}.
\end{eqnarray}
This expression is of the same form as (\ref{F}) and repeating the derivation of Sec.2 we find 
\begin{eqnarray}\label{nhentroy}
 && S=4\zeta(4)   \frac{1}{\pi^{2}}\,\widetilde{V} (T\kappa^{-1})^{3}-3\zeta(3) \left(\frac{1}{8\pi}\,\widetilde{A}\right)(T\kappa^{-1})^{2}\nonumber\\
  &&+ \frac{\zeta(2)}{2\pi^{2}}\left[\int_{\mathcal{S}}d\widetilde{V}\,\frac{(-A_{H}\rho^{2})}{\pi}\left[m^{2}+\left(\xi-\frac{1}{6}\right)\mathcal{R}\right]+\int_{\partial \mathcal{S}}d\widetilde{S}\,
   \frac{\widetilde{K}_{aa}}{3} \right](T\kappa^{-1})\nonumber\\
   && +\frac{1}{16\pi}\int_{\partial \mathcal{S}} d \widetilde{S}\,\left\{\frac{-A_{H}\rho^{2}}{\pi}\left[m^{2}+\left(\xi-\frac{1}{6}\right)\mathcal{R}\right]\right.\nonumber\\
 &&\left.+\frac{1}{96}[8\widetilde{R}_{NaNa}+7\widetilde{K}_{aa}\widetilde{K}_{bb}-10\widetilde{K}_{ab}\widetilde{K}_{ab}]\right\}\log (T\kappa^{-1})+O\left((T\kappa^{-1})^{0}\right).
\end{eqnarray}
Note that all factors of $T$ appear in the combination $T\kappa^{-1}$. So, evaluating the expansion at
$T_{H}=\kappa/2\pi$ we find that all the terms in the expansion become numerical factors. This is the enhancement of the entropy discussed in \cite{Ordonez1}. 

We now give a simple proof of this enhancement in exact entropy without using the high $T$ expansion. Taking the negative $T$ derivative of (\ref{FreeNearH}) we get the following closed expression for the entropy
\begin{eqnarray}
 S =  \frac{1}{4\sqrt{\pi}} \frac{1}{(T\kappa^{-1})^{3}}\sum_{k=1}^{\infty}\int_{0}^{\infty}\frac{du}{u^{5/2}}\,k^{2}\,e^{-\frac{k^{2}}{4u(T\kappa^{-1})^{2}}}e^{-u}\,\text{Tr}\, e^{-u\widetilde{H}}.
\end{eqnarray}
Here, we again observe that in $S$ all factors of $T$ appear in the combination $T\kappa^{-1}$, and at $T=T_{H}$ this combination just equals $(2\pi)^{-1}$. This proves the enhancement property.

A computation similar to the one given for exact metrics shows that (see Appendix A.4) 
 \begin{eqnarray}
   S = A_{H}B_{rs}\frac{1}{\delta^{2}}-C_{rs}\log\frac{A_{H}}{\pi\delta^{2}}.
 \end{eqnarray}
where
\begin{eqnarray}
   B_{rs} = \frac{1}{4\pi^{3}}\left[\frac{\zeta(4)}{\pi^{2}}-\frac{3}{8}\zeta(3)+\frac{2}{3}\zeta(2)-\frac{\pi^{2}}{48}\, \log \frac{1}{2\pi }\right]
\end{eqnarray}
and
\begin{eqnarray}
   C_{rs} =  \frac{\zeta(2)}{8\pi^{3}}\left[m^{2}A_{H}+8\pi\left(\xi-\frac{1}{6}\right)\right].
 \end{eqnarray}
 The coefficient $B_{rs}$ we have here is identical to the one we derived for the exact metrics (see (\ref{BRealScal})), except for the argument of the logarithm, which is now given explicitly by a pure number. In particular, this result verifies, up to the order considered here and up to the complication due to the logarithm, the conjectured equality \cite{Ordonez1} of the high-temperature expansions of the leading horizon divergence for the near-horizon and exact metrics. Computing $B_{rs}$ explicitly, we get  \begin{eqnarray}
   B_{rs} \approx 9.138\times 10^{-3}.
\end{eqnarray}
 
As for the sub-leading horizon divergence, note that the argument of the logarithm is different from the one obtained in the case of exact metrics. To understand this, first note that the variable $\rho$ introduced in (\ref{ro}) is dimensionless. An important consequence of $\rho$ being dimensionless is that in calculating the bulk contributions to the sub-leading logarithmic horizon divergence one may focus only on the lower (brick wall) limits of volume integrals to get a dimensionless argument of the logarithm. As shown in Appendix A.4, this leads to an expression which is proportional to $A_{H}$ and independent of the infrared cut-off. A detailed discussion of this result and its comparison with existing results (\textit e.g. \cite{Fursaev1,Sen} and the review \cite{Page1}) on logarithmic corrections to black hole entropy will be given elsewhere.

On the other hand, if we assume that the region of validity of the near-horizon approximation includes the outer wall at $r_{2}$ (which plays the role of infrared regulator) one may prefer to proceed as in the case of the exact metric and include the contributions of the upper limits of the bulk integrals in calculating the sub-leading horizon divergence. In this case, we recover the form
\begin{eqnarray}
   S = A_{H}B_{rs}\frac{1}{\delta^{2}}+C_{rs}\log\frac{\delta^{2}\kappa}{2(r_{2}-r_{H})}.
 \end{eqnarray}
The difference between the two expressions for the entropy is a finite constant. 

Moreover, comparison of the sub-leading divergence with the sub-leading horizon divergence for the exact metric given in  (\ref{CRSMKAA}), shows that the former has a missing piece, namely the $\zeta(4)\widetilde{V}$ term. As shown in Appendix A.4, due to the specific form of the near-horizon metric, the calculation of $\widetilde{V}$ generates only the leading $O(\delta^{-2})$ divergence. Moreover, since $\mathcal{R}\neq 0$ for the near-horizon metric, while $\mathcal{R}= 0$ in the exact geometry, there is also a discrepancy in the remaining $\zeta(2)$ term of the sub-leading horizon divergence. These are obviously the consequences of retaining only the first term of the Taylor expansion of the exact metric in the derivation of the near-horizon approximation. In order to recover the missing terms one may try to go to higher-orders in the Taylor expansion of the exact metric. For example, it is not difficult to show that inclusion of the sub-leading term in the Taylor expansion produces a term proportional to $\zeta(4)\widetilde{V}$ in the sub-leading divergence. In the present paper we will not consider such higher-order corrections to the exact metric.

\section{Thermodynamics in the Presence of Neumann Walls}
In this section, we will consider the effects of Neumann boundary conditions on the entropy. Since we impose Neumann boundary conditions on the original mode functions $f_{\sigma}$ the rescaled mode functions $\overline{f}_{\sigma}$ will satisfy generalized Neumann boundary conditions
\begin{equation}
\mathcal{B}\phi= (\phi_{;N}+J\phi)\mid_{\partial \mathcal{S}}=0
\end{equation}
We calculate $J$ with a little algebra;
\begin{equation}
\overline{f}_{\sigma}=F^{\frac{1}{2}}f_{\sigma} ,\;\;\; f_{\sigma ;N}\mid_{\partial \mathcal{S}}=0,
\end{equation}
\begin{equation}
\overline{f}_{\sigma ;N}=(\frac{1}{2} F^{-1/2} F_{;N})F^{-1/2}\overline{f}_{\sigma} ,\;\;\; \overline{f}_{\sigma ;N}-\frac{1}{2}\frac{F_{;N}}{F}\overline{f}_{\sigma}=0,
\end{equation}
\begin{equation}
J=-\frac{1}{2}\frac{F_{;N}}{F}.
\end{equation}
For the Neumann problem, the first few heat kernel coefficients are given as
\begin{eqnarray}
  a_{0} &=& \frac{1}{(4\pi)^{3/2}}\int_{\mathcal{S}} dV,\qquad  a_{1/2} = \frac{1}{4(4\pi)}\int_{\partial \mathcal{S}}dS, \nonumber\\
  a_{1} &=& \frac{1}{6(4\pi)^{3/2}}\left[\int_{\mathcal{S}} dV(-6U+R)+2\int_{\partial \mathcal{S}}dS \,(K_{aa}+6J)\right],\nonumber\\
  a_{3/2}&=&\frac{1}{4}\frac{1}{(4\pi)}\frac{1}{96}\int_{\partial \mathcal{S}}dS\,[16(-6U+R)+8 R_{aNaN}+\nonumber\\
  &&+13K_{aa}K_{bb}+2K_{ab}K_{ab}+96JK_{aa}+192J^{2}].
\end{eqnarray}
Comparing these coefficients with the ones for the Dirichlet problem (\ref{a}), we see that here we have some extra terms involving $J$ and some sign changes. With these minor changes, the calculations proceed exactly as in the Dirichlet case. Here we shall report only the final results for the entropy.

The horizon divergence in the entropy again has the general form
\begin{equation}
S_{rs} = B_{rs}\,A_{H}\,\frac{1}{\delta^{2}}
+ C_{rs}(m,\kappa,A_{H},a)\,\log\frac{\kappa\delta^{2}}{2(r_2-r_H)}.
\end{equation}
For all cases, we obtain the coefficient $B_{rs}$ as
\begin{equation}
B_{rs} \;=\; \frac{1}{4\pi^{3}}\!\left[
\frac{\zeta(4)}{\pi^{2}}
+ \frac{3}{8}\zeta(3)
- \frac{4}{3}\zeta(2)
- \frac{5\,\pi^{2}}{48}\,\log\!\left(\frac{\kappa}{2\pi m'}\right)
\right].
\end{equation}
On the other hand, the coefficient $C$ for the Schwarzschild metric is
\begin{equation}
  C_{rs}^{S} = \frac{1}{\pi^{2}}\!\left[
-\frac{\zeta(4)}{\pi^{2}}
+ \frac{\zeta(2)}{8\pi}\,m^{2} A_{H}
\right]  
\end{equation}
for RN metric it is given by
\begin{equation}
  C_{rs}^{RN} = \frac{1}{\pi^{2}}\!\left[
-\frac{\zeta(4)}{\pi^{2}} \,\frac{2r_{+}-3r_{-}}{2r_{+}}
+ \frac{\zeta(2)}{8\pi}\,m^{2} A_{H}
\right],   
\end{equation}
and for the dilatonic metric we find,
\begin{equation}
 C_{rs}^{D} = \frac{1}{\pi^{2}}\!\left[
-\frac{\zeta(4)}{\pi^{2}}\,\frac{(8M-3a)}{8M}
+ \frac{\zeta(2)}{8\pi}\,m^{2} A_{H}
\right].   
\end{equation}
Equivalently, in the compact form that exhibits the dependence on $(m,\kappa,A_{H},a)$ we have,
\begin{equation}
C_{rs}(m,\kappa,A_{H},a)
= \frac{\zeta(4)}{\pi^{4}}\!\left[
\frac{1}{2}
- \frac{3}{2\sqrt{\pi}}\,\kappa \sqrt{A_{H}}\,
\frac{1-a\kappa}{\sqrt{1-2a\kappa}}
\right]
+ \frac{\zeta(2)}{8\pi^{3}}\,m^{2} A_{H}.
\end{equation}
Note that the coefficient $C_{rs}$ is identical to the Dirichlet result. This is not surprising since $C_{rs}$ is determined by the bulk contributions to the heat kernel coefficients, which are the same in both boundary value problems.

In the case of the extreme RN metric we can proceed as in the Dirichlet case (see Sec. 3.4) and find
\begin{equation}
 S= \frac{1}{\pi^{2}}\!\left[
\frac{\zeta(4)}{2\pi^{2}}
+ \frac{\zeta(2)}{8\pi}\,m^{2} A_{H}
\right]\log\frac{\epsilon}{(r_2-r_H)}.
\end{equation}

\section{Conclusion}
In conclusion, we applied 't Hooft's brick wall method \cite{tHooft}, which is an effective tool not only for exploring the black hole horizon but also for renormalizing the gravitational constant \cite{Susskind}, to investigate the thermodynamics of a quantum field around a black hole. In particular, we studied the effect of the brick wall, which is realized by subjecting the quantum field to Dirichlet boundary conditions on a surface (brick wall) just outside the event horizon, on various thermodynamic quantities such as free energy and entropy.

We examined the thermodynamics of a quantum field (real scalar, complex scalar and Dirac) in the Schwarzschild, Reissner-Nordstr\"{o}m, dilaton black hole backgrounds, and also in the near-horizon approximation. We derived high-temperature expansions for the free energy and the entropy of a quantum field in the presence of the brick wall, and studied the boundary and geometric effects on the horizon divergences by examining higher-order terms in the expansion of the entropy. We showed that the coefficient of the leading horizon divergence exhibits a universal behavior characterized by a numerical factor independent of the parameters of the background metric and the quantum field, except for a logarithmic term in surface gravity encountered in the case of the scalar field whose detailed analysis will be given in a future work. We showed that at Hawking temperature, the contributions to this coefficient stemming from different orders of the expansion are enhanced and become comparable. This is in accordance with the near-horizon analysis of \cite{Ordonez1} (see also \cite {Ordonez2, Ordonez3}). In the near-horizon geometry we also gave a concise proof of this enhancement property without using the high-temperature expansion of the entropy. Moreover, we showed that the result for the approximate near-horizon metric is identical with the result for the exact metrics, except for the logarithmic term appearing in the case of the scalar field. This verifies the conjectured equality \cite{Ordonez1} of the leading horizon divergences for exact and near-horizon metrics, at least up to the order considered and up to the aforementioned logarithmic term in the scalar case. 

For the sub-leading horizon divergence, we derived a general formula that covers all cases considered, and gives the coefficient of the divergence as a function of $\kappa, A_{H}, a$ and $m$. Taking an appropriate limit of our result for the non-extremal Reissner-Nordstr\"{o}m, we obtained the horizon divergences for the extremal case and showed that the leading divergence vanishes, leaving the logarithmic term as the sole contribution. This means that the Newton's constant does not get renormalized in the extreme limit. 

We also considered the possibility of using Neumann boundary conditions as an alternative to Dirichlet boundary conditions and obtained results that are qualitatively similar to those found in the Dirichlet problem.

Our derivation of the high-temperature expansion is based on the use of the Mellin transform technique together with the heat kernel expansion to handle harmonic sums representing the free energy and the entropy. These expansions can also be used to analyze the thermodynamics of quantum fields in non-singular gravitational backgrounds.  

 Our work contributes to a deeper understanding of the brick wall regularization in black hole physics, and provides further insights into the complex thermodynamic properties of quantum systems in black hole backgrounds. The results improve our knowledge of intricate connections between gravity, quantum mechanics, and black hole thermodynamics.

\section*{Acknowledgments}

This work is supported by Bo\u{g}azi\c{c}i University BAP Project No. 6942. We thank Prof. Teoman Turgut for useful conversations, and Prof. Carlos Ordonez for bringing the references \cite{Ordonez1,Ordonez2,Ordonez3} to our attention.

\begin{appendices}

\section{Computation of Entropy}
In this appendix, we explicitly evaluate the entropy expression (\ref{sexp}) for the Schwarzschild, RN, and dilaton black holes, as well as for the near-horizon geometry.

\subsection{The Schwarzschild Black Hole}
In the Schwarzschild geometry we have
\begin{equation}
  F=1-\frac{2M}{r},\;\;\;\;\mathcal{R}=0, \;\;\;\; 
\end{equation}
The extrinsic curvature on an $r=\textrm{const.}$ hypersurface is
\begin{equation}
  K_{11}=K_{22}=\frac{3M-r}{r^{2}},\;\;\;K_{12}=0,
\end{equation}
which implies
\begin{equation}
  K_{aa}=2\frac{3M-r}{r^{2}},\;\;\;\;K_{ab}K_{ab}=2\frac{(3M-r)^{2}}{r^{4}}.
\end{equation}
Also
\begin{equation}
  R_{aNaN}=\frac{2M(3M-2r)}{r^{4}}.
\end{equation}
Now using (\ref{volume}) we get the volume of $\mathcal{S}$ as
  \begin{equation}
    V=4\pi\left[\frac{r^{3}}{3}+2Mr^{2}+12M^{2}r+32M^{3}\log(r-2M)-\frac{16M^{4}}{r-2M}\right]_{r_{1}}^{r_{2}}.
  \end{equation}
Therefore
\begin{equation}
  V \asymp  64\pi M^{3}\left(\frac{M}{\epsilon}-2\log \frac{\epsilon}{r_2-r_H}\right).
\end{equation}
Similarly using (\ref{area}) the area of $\partial \mathcal{S}$ is given as
\begin{eqnarray}
  A &=& 4\pi\left(\frac{r^{3}}{r-2M}\right).
\end{eqnarray}
So $A$ diverges as
\begin{equation}
  A\asymp \frac{32\pi M^{3}}{\epsilon}.
\end{equation}
At $O(T)$ the bulk contribution is
\begin{eqnarray}
  \int_{\mathcal{S}}dV\,(-F) m^{2} &=&-4\pi  m^{2}\int_{r_{1}}^{r_{2}}dr\frac{r^{2}}{F} = -4\pi m^{2}\left[(4M^{2}r+Mr^{2}+\frac{r^{3}}{3}+8M^{3}\log(r-2M))\right]_{r_{1}}^{r_{2}}\nonumber\\
   &\asymp  & 32\pi M^{3}m^{2}\log \frac{\epsilon}{r_2-r_H}.
\end{eqnarray}
In the same order, the boundary contribution to the horizon divergence comes only from the integral over $r=r_{1}$ and is seen to be
 \begin{eqnarray}
   \int_{r=r_{1}} dS_{r_{1}}\,\frac{1}{3}\,K_{aa} = \frac{8\pi}{3} \left[\frac{r_{1}(3M-r_{1})}{r_{1}-2M}\right] \asymp \frac{16\pi M^{2}}{3}\frac{1}{\epsilon},
\end{eqnarray}
and
\begin{eqnarray}
&& \frac{1}{16\pi}\int_{\partial \mathcal{S}}dS\,\left\{-F\,m^{2}+\frac{1}{96}[8R_{NaNa}+7K_{aa}K_{bb}-10K_{ab}K_{ab}]\right\} \nonumber\\
&& =\frac{-F\,m^{2}}{4}+ \frac{1}{48}\left[\frac{2M\,(3M-2r)+(3M-r)^{2}}{r\,(r-2M)} \right] \asymp -\frac{M}{96}\, \frac{1}{\epsilon}.
 \end{eqnarray}
If we compare various bulk and boundary terms appearing in the coefficients of the expansion (\ref{sexp}) we get
\begin{eqnarray}
\frac{A}{V}&=&O(M^{-1}),\nonumber\\
\frac{\int_{\partial \mathcal{S}} dS\ \frac{1}{3}\ K_{aa}}{V}&=&O(M^{-2}),\nonumber\\
\frac{\int_B F m^2}{V}&\rightarrow & 0.
\end{eqnarray}
Thus, for generic values of $T$ we see that the boundary contributions are suppressed against the bulk contributions by inverse powers of $M$. However, we must evaluate the entropy at the Hawking temperature $T_{H}$ which is $O(M^{-1})$. In this case, we have for example
\begin{equation}
  \frac{AT^{2}_{H}}{VT^{3}_{H}}=O(M^{0}),
\end{equation}
and the boundary contributions become as important as the bulk terms.

The surface gravity $\kappa$ and the Hawking temperature $T_{H}$ are given by
\begin{equation}
 \kappa=\frac{1}{4 M},\;\;\;\;\;\;\;\; T_{H}=\frac{\kappa}{2\pi}=\frac{1}{8\pi M},
\end{equation}
We relate the cut-off $\epsilon$ to the proper length cut-off $\delta$ using (\ref{deleps})
 \begin{equation}
\epsilon\asymp T_{H}\pi\delta^2=\frac{\delta^2}{8 M},
 \end{equation}
 and note that the horizon area is
 \begin{equation}
   A_{H}=16\pi M^{2}.
 \end{equation}
 Then (\ref{sexp}) is explicitly given as
 \begin{eqnarray}
   S &=& \frac{A_{H}}{4\pi^{3}}\left[\frac{\zeta(4)}{\pi^{2}}-\frac{3}{8}\zeta(3)+\frac{2}{3}\zeta(2)-\frac{\pi^{2}}{48}\, \log \frac{\kappa}{2\pi m'}\right]\frac{1}{\delta^{2}}\nonumber\\
 &+&\frac{1}{\pi^{2}}\left[ -\frac{\zeta(4)}{\pi^{2}}+\frac{\zeta(2)}{8\pi}m^{2}\,A_{H}\right] \log\frac{\kappa\delta^{2}}{2(r_2-r_H)}.
 \end{eqnarray}

\subsection{Reissner-Nordstr\"{o}m Black Hole}\label{RNapp}
For the RN background we have
\begin{equation}
  F(r)=\left(1-\frac{r_{-}}{r} \right) \left( 1- \frac{r_{+}}{r}\right),\;\;\;\;\mathcal{R}=0. 
\end{equation}
On an $r=\textrm{const.}$ hypersurface
\begin{equation}
 K_{11}=K_{22}=\frac{3r(r_{+}+r_{-})-2r^{2}-4r_{+}r_{-}}{2r^{3}},\;\;\;K_{12}=0,
\end{equation}
which implies
\begin{equation}
  K_{aa}=\frac{-2r^{2}+3(r_{+}+r_{-})r-4r_{+}r_{-}}{r^{3}},\;\;\;\;K_{ab}K_{ab}=2\left[ \frac{3r(r_{+}+r_{-})-2r^{2}-4r_{+}r_{-}}{2r^{3}}\right] ^{2}.
\end{equation}
We also have
\begin{equation}
  R_{aNaN}=\frac{8r_{+}^{2}r_{-}^{2}-12r_{+}r_{-}(r_{+}+r_{-})r+3(r_{+}^{2}+6r_{+}r_{-}+r_{-}^{2})r^{2}-4(r_{+}+r_{-})r^{3}}{2r^{6}}.
\end{equation}
Thus, using (\ref{volume}) we get the volume of the spherical shell as
\begin{eqnarray}
V &=& 4\pi \left[  \frac{r_{-}^{6}}{(r_{-}-r_{+})^{2} (r_{-}-r)} + \frac{r_{+}^{6}}{(r_{-}-r_{+})^{2} (r_{+}-r)} + \left( 3r_{-}^{2}+4r_{-} r_{+} +3r_{+}^{2}\right) r \right.   \nonumber\\
 &+& \left( r_{-}+r_{+}\right) r^{2} + \frac{r^{3}}{3} + \frac{2r_{-}^{5}(2r_{-}-3r_{+}) \log(r-r_{-})}{(r_{-}-r_{+})^{3}}\nonumber\\
 &+& \left. \frac{2(3r_{-}-2r_{+}) r_{+}^{5} \log(r-r_{+})}{(r_{-}-r_{+})^{3}}\right]_{r_{1}}^{r_{2}}.\nonumber \\
 &&
\end{eqnarray}
Hence
\begin{equation}
V \asymp 4\pi\frac{r_{+}^{5}}{(r_{-}-r_{+})^{2}} \left[ \frac{r_{+}}{\epsilon}+ \frac{3r_{-}-2r_{+}}{r_{+}-r_{-}} 2\log \frac{\epsilon}{r_2-r_H} \right].
\end{equation}
Similarly, using (\ref{area}) we get
\begin{equation}
A \asymp  \frac{4\pi r_{+}^{4}}{r_{+}-r_{-}}\frac{1}{\epsilon}.
\end{equation}
On the other hand
\begin{eqnarray}
\int_\mathcal{S} (-F)\,m^{2} &=& -4\pi m^{2} \int_{r_{1}}^{r_{2}} dr \frac{r^{2}}{F} \nonumber\\
&=& -4\pi m^{2}\left [ (r_{+}^{2}+r_{-}^{2} +r_{+} r_{-}) r+ \frac{1}{2} (r_{+}+r_{-})r^{2} +\frac{r^{3}}{3} \right.\nonumber\\
 &+& \left. \frac{r_{-}^{4}}{r_{+}-r_{-}} \log (r-r_{-})+\frac{r_{+}^{4}}{r_{+}-r_{-}} \log(r-r_{+}) \right] _{r_{1}}^{r_{2}} \nonumber\\
&\asymp & \frac{4\pi\,m^{2} r_{+}^{4}}{r_{+}-r_{-}}  \log \frac{\epsilon}{r_2-r_H}.
\end{eqnarray}
Finally, using (\ref{extrinsic}), the surface integral of extrinsic curvature over the brick wall is given as
\begin{equation}
\int_{r=r_{1}}dS_{r}\, K_{aa} = -4\pi \frac{2r_{1}^{3}-3r_{1}^{2}(r_{+}+r_{-})+4r_{1}r_{+}r_{-}}{(r_{1}-r_{-})(r_{1}-r_{+})}.
\end{equation}
Thus
\begin{equation}
\int_{\partial \mathcal{S}} dS\ \frac{1}{3}\ K_{aa} \asymp \frac{4\pi}{3}  \frac{r_{+}^{2}}{\epsilon},
\end{equation}
and
\begin{eqnarray}
&& \frac{1}{16\pi}\int_{\partial \mathcal{S}}dS\,\left\{-F\,m^{2}+
 \frac{1}{96}[8R_{NaNa}+7K_{aa}K_{bb}-10K_{ab}K_{ab}]\right\} \nonumber\\
 && \asymp -\frac{(r_{+}-r_{-})}{192} \, \frac{1}{\epsilon}.
 \end{eqnarray}
The surface gravity and the Hawking temperature are given by
\begin{equation}
 \kappa=\frac{r_{+}-r_{-}}{2 r_{+}^{2}},\;\;\;\;\;\;\;\; T_{H}=\frac{\kappa}{2\pi}=\frac{r_{+}-r_{-}}{4\pi r_{+}^{2}},
\end{equation}
and
\begin{equation}
  A_{H}=4\pi \, r_{+}^{2}.
\end{equation}
Using these we relate the cut-off $\epsilon$ to the proper length cut-off $\delta$ using (\ref{deleps}) 
\begin{equation}
  \epsilon \asymp \frac{\kappa\delta^{2}}{2}=\frac{(r_{+}-r_{-})}{4r_{+}^{2}}\delta^{2},
\end{equation}
we obtain
\begin{eqnarray}
  S &=& \frac{A_{H}}{4\pi^{3}}\left[\frac{\zeta(4)}{\pi^{2}}-\frac{3}{8}\zeta(3)+\frac{2}{3}\zeta(2)-\frac{\pi^{2}}{48}\, \log \frac{\kappa}{2\pi m'}\right] \frac{1}{\delta^{2}}\nonumber\\
  &+&\frac{1}{\pi^{2}}\left[-\frac{\zeta(4)}{\pi^{2}} \, \frac{2r_{+}-3r_{-}}{2r_{+}}+\frac{\zeta(2)}{8\pi}m^{2}\, A_{H} \right] \log\left(\frac{\kappa\delta^{2}}{2(r_2-r_H)} \right).
\end{eqnarray}
Note that the coefficient of the $\delta^{-2}$ term is the same as in the Schwarzschild case.

Following \cite{Alwis0, Alwis}, the extreme RN case is studied by taking $ r_{-} \rightarrow r_{+}$ limit of the results we found for the non-extremal metric. In this limit 
\begin{equation}
 \kappa\rightarrow 0,\;\;\;\;\;\delta=\sqrt{\frac{2\epsilon}{\kappa}}\rightarrow \infty
\end{equation}
and
\begin{equation}
 \frac{\kappa\delta^2}{2}\rightarrow \epsilon.
\end{equation}
Hence,
\begin{eqnarray}
  S =\frac{1}{\pi^{2}}\left[\frac{\zeta(4)}{\pi^{2}} \, \frac{1}{2}+\frac{\zeta(2)}{8\pi}m^{2}\, A_{H}\right] \log\frac{\epsilon}{r_2-r_H}.
\end{eqnarray}

\subsection{Dilaton Black Hole}\label{DBapp}
For the dilaton black hole metric
\begin{equation}
  F(r)=\left( 1-\frac{2M}{r}\right),\;\;\;\;\mathcal{R}=\frac{a^{2}(r-2M)}{2(r-a)^{2} r^{3}},
\end{equation}
On an $r=\textrm{const.}$ hypersurface we have
\begin{equation}
K_{11}=K_{22}=\frac{(-4 a M + a r + 6 M r - 2 r^2)}{2 r^2 (r - a)},\;\;\;K_{12}=0,
\end{equation}
and consequently
\begin{equation}
  K_{aa}=\frac{-2r^{2}+(6M+a)r-4aM}{r^{2}(r-a)},\;\;\;\;K_{ab}K_{ab}=2\left[ \frac{(-4 a M + a r + 6 M r - 2 r^2)}{2 r^2 (r - a)}\right] ^{2}.
\end{equation}
We also have
\begin{equation}
R_{aNaN}=\frac{4 M (3 M - 2 r) r^2 + 8 a M r (-3 M + 2 r) +
 a^2 (16 M^2 - 12 M r + r^2)}{2 (a - r)^2 r^4}.
\end{equation}
So we get
\begin{equation}
V \asymp 4\pi \left[ 8M^{3}(2M-a)\frac{1}{\epsilon} + 4M^{2}(3a-8M) \log \frac{\epsilon}{r_2-r_H} \right],
\end{equation}
\begin{equation}
A \asymp 4\pi \left[ 4 M^{2}(2M-a)\frac{1}{\epsilon} \right],
\end{equation}
\begin{equation}
\int_\mathcal{S} dV\,(-F)\left[ m^{2}+\left( \xi-\frac{1}{6}\right) \mathcal{R} \right] \asymp 16\pi m^{2}M^{2}(2M-a)\log\epsilon.
\end{equation}
Integrating over the boundary
\begin{equation}
\int_{\partial \mathcal{S}} dS \frac{1}{3}K_{aa} \asymp \frac{8\pi}{3}   M (2M-a) \frac{1}{\epsilon},
\end{equation}
and
\begin{eqnarray}
 &&\int_{\partial \mathcal{S}}\,\frac{dS}{16\pi}\left\{-F\left[m^{2}+\left(\xi-\frac{1}{6}\right)\mathcal{R
 }\right]+\frac{1}{96}[8R_{NaNa}+7K_{aa}K_{bb}-10K_{ab}K_{ab}]\right\} \nonumber\\
 &&\asymp  -\frac{(2M-a)}{192} \,\frac{1}{\epsilon}.
 \end{eqnarray}
Plugging the above results into (\ref{sexp}) together with
\begin{equation}
A_{H}=4\pi\, 2M(2M-a),\;\;\;\;\kappa=\frac{1}{4 M},\;\;\;\; T_{H}=\frac{\kappa}{2\pi}=\frac{1}{8\pi M},\;\;\;\; 
\epsilon=\frac{\delta^{2}}{8M}\;\;
\end{equation}
we get
\begin{eqnarray}\label{dilatonS}
   S &=& \frac{A_{H}}{4\pi^{3}}\left[\frac{\zeta(4)}{\pi^{2}}-\frac{3}{8}\zeta(3)+\frac{2}{3}\zeta(2)-\frac{\pi^{2}}{48}\, \log \frac{\kappa}{2\pi m'
   }\right]\frac{1}{\delta^{2}}\nonumber\\
 &+&\frac{1}{\pi^{2}}\left[ -\frac{\zeta(4)}{\pi^{2}}\frac{(8M-3a)}{8M}+\frac{\zeta(2)}{8\pi}m^{2}\,A_{H}\right] \log\frac{\kappa\delta^{2}}{2(r_2-r_H)}.
 \end{eqnarray}
Note that the coefficient of the $\delta^{-2}$ term is the same as in the Schwarzschild and RN cases.

\subsection{Entropy in Near-Horizon Geometry}

From the simplified optical metric (see (\ref{metilde})), we have
\begin{equation}
d\tilde{s}^2 = \frac{d\rho^2}{\rho^2} + \frac{1}{4\rho^2}\left(d\theta^2 + \sin^2\theta\, d\phi^2\right).
\end{equation}
Direct computation gives the following divergence in volume as follows:
\begin{equation}
  \widetilde{V} = 4\pi \int_{\rho(\epsilon)}^{L} \frac{d\rho}{4\rho^{3}} 
  \asymp \frac{\pi}{2\rho^{2}(\epsilon)} = \frac{A_{H}}{2\delta^{2}},
\end{equation}
and similarly,
\begin{equation}
  \widetilde{A} \asymp \frac{\pi}{\rho^{2}(\epsilon)} = \frac{A_{H}}{\delta^{2}}.
\end{equation}
Here, as noted in (\ref{pos}),
\begin{equation}\label{pos}
    \rho(\epsilon)=\sqrt{\frac{2\pi\epsilon}{\kappa A_{H}}}=\sqrt{\frac{\pi}{A_{H}}}\delta.
\end{equation}

If we choose the following orthonormal basis adapted to $\rho=\text{const.}$ surfaces,
\begin{equation}
  \widetilde{N} = \rho\,\partial_{\rho}, \quad
  \widetilde{E}_{1} = 2\rho\,\partial_{\theta}, \quad
  \widetilde{E}_{2} = \frac{2\rho}{\sin\theta}\,\partial_{\phi},
\end{equation}
then, from (\ref{extrinsicmat}), the extrinsic curvature of a $\rho=\text{const.}$ surface is calculated to be
\begin{equation}
   \widetilde{K}_{ab} = \delta_{ab}.
\end{equation}
We also have
\begin{equation}
   \widetilde{R} = -6 + 8\rho^{2}, \qquad \widetilde{R}_{aNaN} = -2.
\end{equation}
From the near-horizon metric, the scalar curvature reads
\begin{equation}
    \mathcal{R} = \frac{2}{r_{H}(r_{H}-a)} = \frac{8\pi}{A_{H}}.
\end{equation}
The bulk integral becomes
\begin{equation}
-\int_{\mathcal{S}} d\widetilde{V}\,\frac{A_{H}\rho^{2}}{\pi}
\left[m^{2} + \left(\xi - \frac{1}{6}\right)\mathcal{R}\right]
\asymp -\frac{1}{2}\left[m^{2}A_{H} + 8\pi\left(\xi - \frac{1}{6}\right)\right]
\log\frac{A_{H}}{\pi\delta^{2}}.
\end{equation}
Note that here we do not include the finite contribution coming from the upper limit of the integral in the resulting singular term. Since we work with the dimensionless variable $\rho$, the argument of the logarithm is already dimensionless; there is no need for the upper limit of the integral to obtain a dimensionless argument of the logarithm. On the other hand, if we include the contribution of the upper limit we get
\begin{equation}
-\int_{\mathcal{S}} d\widetilde{V}\,\frac{A_{H}\rho^{2}}{\pi}
\left[m^{2} + \left(\xi - \frac{1}{6}\right)\mathcal{R}\right]
\asymp \frac{1}{2}\left[m^{2}A_{H} + 8\pi\left(\xi - \frac{1}{6}\right)\right]
\log\frac{\kappa\delta^{2}}{2(r_2-r_H)}.
\end{equation}

Integrating over the boundary, we find
\begin{equation}
\int_{\partial \mathcal{S}} d\widetilde{S}\,\frac{\widetilde{K}_{aa}}{3} 
\asymp -\frac{2 A_H}{3\delta^2},
\end{equation}
and
\begin{align}
 &\frac{1}{16\pi} \int_{\partial \mathcal{S}} d\widetilde{S}\,
 \left\{
 -\frac{A_{H}\rho^{2}}{\pi}\left[m^{2} + \left(\xi - \frac{1}{6}\right)\mathcal{R}\right]
 + \frac{1}{96}\left[8\widetilde{R}_{NaNa} + 7\widetilde{K}_{aa}\widetilde{K}_{bb} 
 - 10\widetilde{K}_{ab}\widetilde{K}_{ab}\right]
 \right\} \nonumber\\
 &\asymp -\frac{A_H}{4\pi^3}
 \left[\frac{1}{48}\log\left(\frac{\kappa}{2\pi}\right)\right]
 \frac{1}{\delta^2}.
\end{align}
Plugging the above results into (\ref{nhentroy}), together with
\begin{equation}
A_{H} = 4\pi r_H(r_H - a), \qquad T_{H} = \frac{\kappa}{2\pi},
\end{equation}
we obtain
\begin{align}
   S &= \frac{A_{H}}{4\pi^{3}}
   \left[\frac{\zeta(4)}{\pi^{2}} - \frac{3}{8}\zeta(3) + \frac{2}{3}\zeta(2) 
   - \frac{\pi^{2}}{48}\log \frac{1}{2\pi}\right]\frac{1}{\delta^{2}} \nonumber\\
   &\quad -\frac{\zeta(2)}{8\pi^{3}}
   \left[m^{2}A_{H} + 8\pi\left(\xi - \frac{1}{6}\right)\right]
   \log\frac{A_{H}}{\pi\delta^{2}}.
\end{align}

\end{appendices}


\begin{thebibliography}{44}
	
\bibitem{Hawking1} S. Hawking, Comm. Math. Phys. \textbf{43}, 199 (1975).

\bibitem{Hawking2} J. B. Hartle and S. Hawking,  Phys. Rev. \textbf{D13}, 2188 (1976).

\bibitem{Bekenstein} J. Bekenstein,  Phys. Rev. \textbf{D7}, 2333 (1974).

\bibitem{tHooft} G. 't Hooft,  Nucl. Phys. \textbf{B256}, 727 (1985).

\bibitem{Davies} N. D. Birrell and P. C. W. Davies, \textit{Quantum Fields in Curved Space}, Cambridge University Press (1984).

\bibitem{Vega} H. J. deVega and N. Sanchez, Nucl. Phys. \textbf{B299}, 818 (1988).

\bibitem{Susskind} L. Susskind and J. Uglum, Phys. Rev.  \textbf{D50}, 2700 (1994).

\bibitem{Kabat} D. Kabat and M. J. Strassler, Phys. Lett.  \textbf{B329}, 46 (1994).

\bibitem{Solodukhin} S. N. Solodukhin, Phys. Rev. \textbf{D51}, 618 (1995).

\bibitem{Barbon} J. L. F. Barbon, Phys. Rev. \textbf{D50}, 2712 (1994).

\bibitem{Emperan} R. Emperan, Phys. Rev. \textbf{D51}, 5716 (1995).

\bibitem{Alwis0} S. P. de Alwis and N. Ohta, hep-th/9412027.

\bibitem{Alwis} S. P. de Alwis and N. Ohta, Phys. Rev. \textbf{D52}, 3529 (1995).

\bibitem{Solodukhin1} S. N. Solodukhin, Phys. Rev. \textbf{D51}, 609 (1995).



\bibitem{Fursaev} D. V. Fursaev, Mod. Phys. Lett. \textbf{A10}, 649 (1995).

\bibitem{Belgiorno} F. Belgiorno and M. Martellini, Phys. Rev. \textbf{D53}, 7073 (1996).

\bibitem{Akant} L. Akant, E. Ertugrul, Y. Gul and O. T. Turgut, J. Math. Phys \textbf{56}, 7, (2015) .




\bibitem{Dowker1} J. S. Dowker and G. Kennedy, J. Phys. A: Math. Gen. \textbf{11}, 895 (1978).

\bibitem{Actor} A. Actor, J. Phys.  \textbf{A20}, 5351 (1987).

\bibitem{Dowker2} J. S. Dowker and J. P. Schofield, Phys. Rev.  \textbf{D38}, 3327 (1988).

\bibitem{Dowker3} J. S. Dowker and J. P. Schofield, Nucl. Phys.  \textbf{B327}, 267 (1989).

\bibitem{Kirsten1} K. Kirsten, J. Phys.  \textbf{A24}, 3281 (1991).

\bibitem{Kirsten2} K. Kirsten, Class. Quantum Grav.  \textbf{8}, 2239 (1991).

\bibitem{T1} D. J. Toms, Phys. Rev. Lett. \textbf{69}, 1152 (1992).

\bibitem{T2} D. J. Toms, Phys. Rev.  \textbf{D47}, 2483 (1993).

\bibitem{Ordonez1} H. E. Camblong and C. R. Ordonez, JHEP \textbf{0712}, 099 (2007).

\bibitem{Flajolet} P. Flajolet, X. Gourdon and P. Dumas, Theoretical Computer Science  \textbf{144}, 3 (1995).

\bibitem{gibbons} G. W. Gibbons and M. J. Perry, Proc. Roy. Soc. London  \textbf{A358}, 467 (1978).

\bibitem{FrolovFursaev} V. P. Frolov, D. V. Fursaev and A. I. Zelnikov, Phys. Rev.  \textbf{D54}, 2711 (1996).








\bibitem{Vanzo1} G. Cognola, L. Vanzo and S. Zerbini, Class. Quant. Grav.  \textbf{12}, 1927 (1995).

\bibitem{Vanzo2} G. Cognola, L. Vanzo and S. Zerbini, Phys. Rev.   \textbf{D52}, 4548 (1995).


\bibitem{Ghosh1} A. Ghosh and P. Mitra, Phys. Rev. Lett.   \textbf{73}, 2521 (1994).

\bibitem{Ghosh2} A. Ghosh and P. Mitra, Phys. Lett.   \textbf{B357}, 295 (1995).

\bibitem{Ghosh3} K. Ghosh, Nucl. Phys.   \textbf{B814}, 212 (2009).

\bibitem{Ghosh4} K. Gosh, J. Phys. Soc. Jpn. 85, 014101 (2016).

\bibitem{Ordonez2} H. E. Camblong and C. R. Ordonez, Phys. Rev. \textbf{D71}, 104029 (2005).

\bibitem{Ordonez2.5} H. E. Camblong and C. R. Ordonez, Phys. Rev. \textbf{D71}, 124040 (2005).

\bibitem{Ordonez3} H. E. Camblong and C. R. Ordonez, Class. Quant. Grav. \textbf{30}, 175007 (2013).

\bibitem{Sarkar} S. Sarkar, S. Shankaranarayanan, L. Sriramkumar, Phys. Rev. \textbf{D78}, 024003 (2008).

\bibitem{Kim} W. Kim, S. Kulkarni, EPJ \textbf{C73} 2398 (2013).

\bibitem{Emelyanov} S. Emelyanov,  arXiv:1504.05164.

\bibitem{FrolovNovikov} V. P. Frolov and I. Novikov, Phys. Rev. \textbf{D48}, 4545 (1993).

\bibitem{Barvinsky} A. D. Barvinsky, V. P. Frolov and A I. Zelnikov, Phys. Rev. \textbf{D51}, 1741 (1995).

\bibitem{Solodukhin2} S. N. Solodukhin, Living Rev. Rel.  \textbf{14}, 8 (2011).

\bibitem{Thorne} K. S. Thorne, D. A. MacDonald and R. H. Price, \textit{Black Holes: The Membrane Paradigm}, Yale University Press (1986).



\bibitem{JorLang} J. Jorgenson and S Lang, \textit{Basic Analysis of Regularized Series and Products}, Springer-Verlag (1993).

\bibitem{Gilkey} T. P. Branson and P. B. Gilkey, Comm. Partial Differential Equations \textbf{15}
245 (1990).

\bibitem{Vas} D. V. Vassilevich, Phys. Rept. \textbf{388}
279 (2003).

\bibitem{GibMa} G. Gibbons and K. Maeda, Nucl. Phys. \textbf{298}, 741 (1988).

\bibitem{Garfinkle} D. Garfinkle and A. Strominger, Phys. Rev. \textbf{D43}, 3140; \textbf{D45}, 3888 (E) (1992).


\bibitem{Qiu} W. Qiu, J. Xu, R. Su, B. Wang, Mod. Phys. Lett. \textbf{A16}, 489 (2001). 

\bibitem{Fursaev1} D. V. Fursaev, Phys. Rev. \textbf{D51}, R5352  (1992).

\bibitem{Sen} A. Sen,  JHEP 1304 156 (2013).

\bibitem{Page1} D. N. Page New J. Phys. \textbf{7}, 203 (2005).


\end{thebibliography}
\end{document}